\documentclass[sigplan,10pt]{acmart}

\usepackage{algorithm}%
\usepackage{algorithmicx}%

\usepackage{tabularx}
\usepackage{multirow}
\usepackage{algpseudocode}%

\newcolumntype{C}[1]{>{\centering\arraybackslash\fontsize{#1}{#1*1.2}\selectfont}X}
\algnewcommand{\LineComment}[1]{\State \textcolor{blue}{// #1}}

\newtheorem{theorem}{Theorem}

\AtBeginDocument{%
  }

\setcopyright{none}
\copyrightyear{2027}
\acmYear{2027}
\acmDOI{XXXXXXX.XXXXXXX}
\acmConference[PPoPP '27]{32nd ACM SIGPLAN Annual Symposium on Principles and Practice of Parallel Programming}{March 20--24,
  2027}{Salt Lake City, Utah, USA}

\acmSubmissionID{48}

\begin{document}

\title{Ascend-SpMM: Accelerating Sparse Matrix-Dense Matrix Multiplication on Ascend NPUs}

\author{Jiahe Wang}
\affiliation{%
  \institution{Beijing University of Posts and Telecommunications}
  \city{Beijing}
  \country{China}
}
\email{meaningful@bupt.edu.cn}

\author{Yingxia Shao}
\authornote{Corresponding author.}
\affiliation{%
  \institution{Beijing University of Posts and Telecommunications}
  \city{Beijing}
  \country{China}
}
\email{shaoyx@bupt.edu.cn}

\author{Shuhan Wang}
\affiliation{%
  \institution{Beijing University of Posts and Telecommunications}
  \city{Beijing}
  \country{China}
}
\email{wangshuhan@bupt.edu.cn}

\author{Qingbo Ge}
\affiliation{%
  \institution{Beijing University of Posts and Telecommunications}
  \city{Beijing}
  \country{China}
}
\email{geqingbo@bupt.edu.cn}

\author{Hongzheng Li}
\affiliation{%
  \institution{Beijing University of Posts and Telecommunications}
  \city{Beijing}
  \country{China}
}
\email{Ethan\_Lee@bupt.edu.cn}

\author{Yucheng Zhang}
\authornotemark[1]
\affiliation{%
  \institution{Huawei Co., Ltd}
  \city{Beijing} 
  \country{China}
}
\email{zhangyucheng23@huawei.com}

\renewcommand{\shortauthors}{Jiahe Wang et al.}

\begin{abstract}
  Sparse Matrix-Dense Matrix Multiplication (SpMM) is a fundamental kernel in graph neural networks and scientific computing, yet its optimization on Ascend NPUs remains largely unexplored. The physically decoupled Cube/Vector architecture and single-threaded scalar instruction dispatch of modern NPUs introduce unique challenges. We present Ascend-SpMM, a novel and systematic framework tailored to the physically decoupled architecture of Ascend NPUs. We introduce a memory-centric cost model that quantifies hardware affinity and guides a conflict-bounded, density-aware task-mapping strategy. The strategy maps dense regions to the high-throughput Cube path and sparse residuals to the flexible Vector path, with a theoretical bound on cross-processor conflicts. To execute these workloads with high efficiency, we further develop highly customized sub-kernels. These kernels leverage hierarchy-aware pipelining and round-robin ring buffer-based implicit synchronization to effectively mask memory latency and reduce instruction pipeline stalls. Extensive experiments demonstrate that Ascend-SpMM achieves a geometric-mean speedup of 12.70--20.55× over the strongest evaluated NPU baseline for each matrix, accelerates end-to-end GNN training by up to 6.8×, and delivers performance competitive with state-of-the-art GPU implementations. The source code of Ascend-SpMM is publicly available at \url{https://github.com/IDMG-Lab/Ascend-SpMM}.

\end{abstract}





\maketitle

\section{Introduction}\label{sec1}

\begin{sloppypar}
Sparse Matrix-Dense Matrix Multiplication (SpMM) is a fundamental operator in modern computing, serving as the computational backbone for Graph Neural Networks (GNNs)\cite{gnnsurvey,GCN,GIN}, scientific computing\cite{sci1,sci2}, sparse attention\cite{GAT,seerattn}, and modern sparsity-aware Large Language Model (LLM) inference\cite{llm1,llm2,llm3,llm4}. As the scale and complexity of these applications continue to grow, accelerating SpMM has become increasingly critical. While Graphics Processing Units (GPUs) have traditionally dominated the acceleration of dense and sparse linear algebra, domain-specific Neural Processing Units (NPUs)---represented by the Ascend AI processor\cite{Ascend}---have emerged as a powerful computing force in data centers, offering massive computational throughput and high memory bandwidth tailored for deep learning.
\end{sloppypar}


However, high-performance SpMM on Ascend NPUs cannot be achieved by simply porting existing state-of-the-art GPU kernels, primarily due to two fundamental architectural mismatches. The first mismatch lies in the physically decoupled architecture and its independent bandwidth allocation. Modern NPUs isolate dense matrix operations into AI Cube (AIC) units and fine-grained operations into AI Vector (AIV) units, each possessing independent memory access paths and bandwidth capacities. Since SpMM is a memory-bound operator, execution methods that utilize only one type of hardware resource are inherently insufficient. Cube-centric approaches like Moustafa’s \cite{moustafa} force irregular data into the AIC through zero-padding, saturating the AIC bandwidth with invalid zeros while the AIV's bandwidth remains unused. While NeutronSparse \cite{neutronsparse} recently attempted to address this via heterogeneous workload coordination, its heuristic-based task mapping lacks rigorous theoretical foundations, leading to sub-optimal generalizability across highly diverse density patterns.

The second mismatch stems from the single-threaded execution model of NPUs. Unlike GPUs that rely on hardware-managed warp schedulers and implicit caching to transparently hide memory latency, NPUs employ a single-threaded Scalar Unit for explicit instruction dispatching. However, different hardware units (i.e., memory access units and compute units) possess independent instruction pipelines, relying on explicit synchronization among them to achieve latency hiding. This execution model introduces distinct architectural challenges: on the AIC, the memory hierarchy is highly complex, demanding explicit data and instruction orchestration across deep, multi-level memory buffers; on the AIV, while the architecture is relatively simple, throughput is severely bottlenecked by the limited number of hardware synchronization flags. Without carefully leveraging these NPU-specific architectural features, instruction streams can become serialized, leaving hardware resources severely underutilized.

To address the aforementioned mismatches, we propose Ascend-SpMM, a systematic framework explicitly tailored for the decoupled architecture and single-threaded execution model of Ascend NPUs. First, to exploit the physically decoupled architecture, we introduce a conflict-bounded density-aware heterogeneous task mapping strategy. Guided by a rigorous memory-centric cost model, our approach achieves load balancing with a theoretical guarantee by mapping structured dense workload to the high-throughput AICs while delegating highly irregular workload to the flexible AIVs with a bounded number of atomic-write conflicts. Second, to overcome the absence of hardware warp schedulers and eliminate pipeline stalls, we develop highly customized sub-kernels featuring hierarchy-aware pipelining and implicit synchronization. By scheduling the scalar-dispatched instruction sequence across the explicit multi-level memory hierarchy, and transforming hardware-limited synchronization flags into spatial ring-buffer implicit synchronization, our design achieves increased overlap between memory access and computation. 

In summary, our main contributions are as follows:

\begin{itemize}
    \item We identify the main architectural bottlenecks of Ascend NPUs caused by physically decoupled heterogeneous resources and strict scalar instruction dispatching.
    \item We propose Ascend-SpMM, a systematic hardware-aware software framework explicitly tailored to the NPU's decoupled architecture, effectively transforming hardware limitations into system optimization opportunities.
    \item We propose a conflict-bounded density-aware heterogeneous task mapping strategy. Under a continuous relaxation, we show that the optimal assignment follows a density-threshold structure with limited atomic conflicts.
    \item We design highly optimized NPU-native SpMM sub-kernels featuring hierarchy-aware pipelining and implicit synchronization, which increase bandwidth utilization and overlap between memory access and computation without relying on hardware multi-threading.
    \item Extensive evaluations on diverse real-world datasets demonstrate that Ascend-SpMM achieves a geometric-mean speedup of up to 20.55$\times$ over the strongest evaluated NPU baseline. Furthermore, it delivers highly competitive performance relative to the state-of-the-art GPU implementations.
\end{itemize}

\section{Background and Motivations}\label{sec2}

\begin{figure}[tb]
    \centering
    \includegraphics[width=\linewidth]{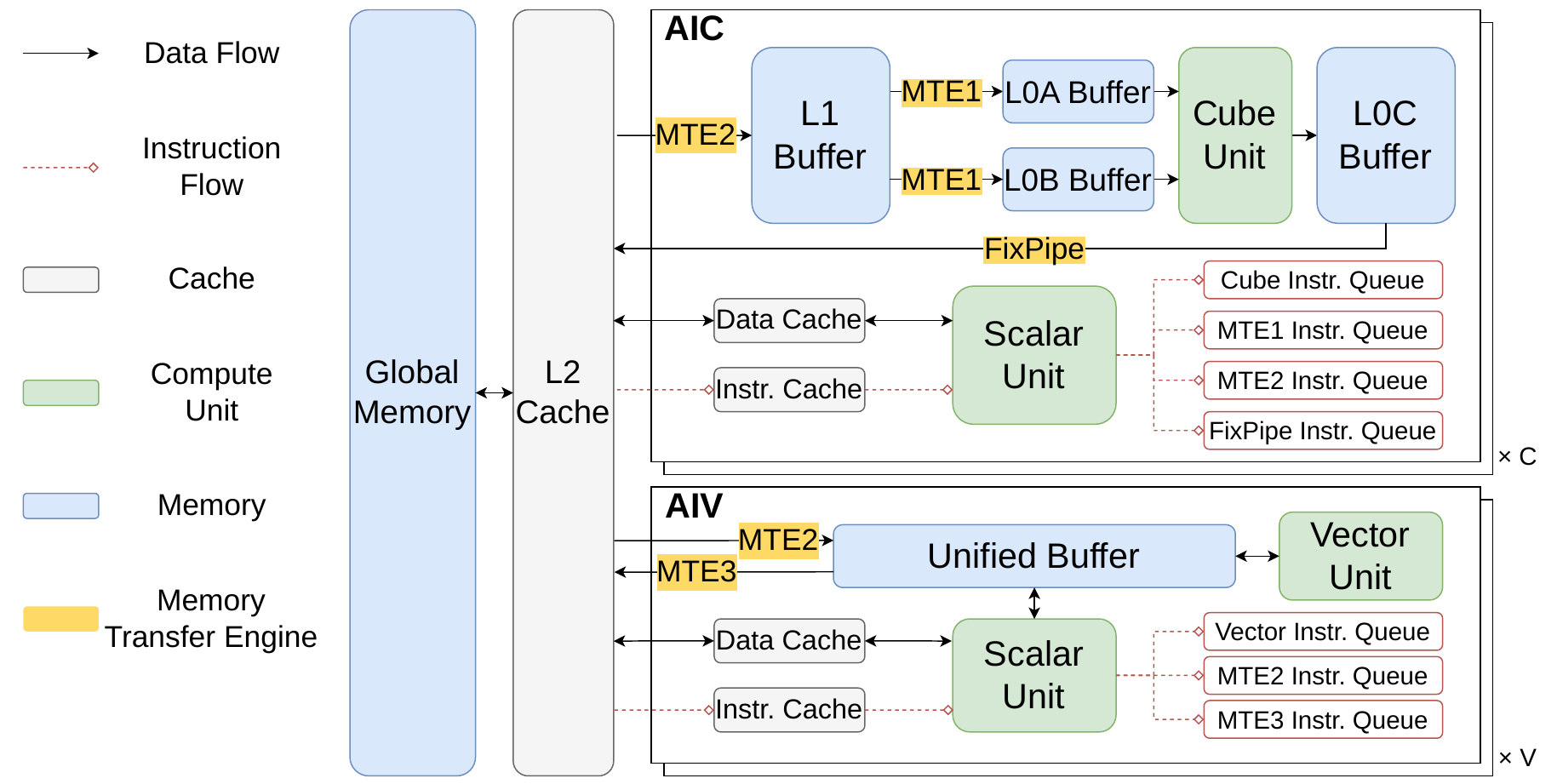}
    \caption{\small The Ascend NPU Decoupled Architecture.}
    \label{fig1}
\Description{} \end{figure}

\subsection{Ascend NPU Architecture}\label{subsec21}

Ascend is a domain-specific NPU designed for deep learning workloads\cite{Ascend, davinci}. As shown in Figure~\ref{fig1}, it adopts a physically decoupled architecture: the AI Cube (AIC) for matrix operations and the AI Vector (AIV) for vector operations are separate cores with independent computing units, on-chip memory, and memory access engines\cite{NPUarch1,NPUarch2}. 

\textbf{Compute Units.} An NPU contains three types of compute units: the Scalar Unit, which is responsible for scalar data processing and program control flow; the Vector Unit, which performs vector operations in the Single Instruction Multiple Data (SIMD) paradigm; and the Cube Unit, which supports FP16 and FP32 matrix multiplications.

\textbf{Memory Hierarchy.} The memory hierarchy comprises software-managed Global Memory (GM), Unified Buffer (UB), and L0/L1 buffers exposing data placement to developers, and hardware-managed caches. Data movement is orchestrated by Memory Transfer Engines (MTEs), each serving a specific memory level. Transfers within the same MTE are executed sequentially, while transfers across different MTEs can proceed in parallel.

\textbf{Execution Model.} Ascend employs an explicit instruction pipelining mechanism to hide latency. Both compute operations and memory transfers are abstracted as instructions and dispatched by the Scalar Unit to different execution queues corresponding to the Scalar, Vector, Cube, and MTE units. Instructions within the same queue are executed in FIFO order, whereas instructions in different queues can be executed concurrently.

\subsection{Sparse Matrix-Dense Matrix Multiplication}\label{subsec22}

SpMM is mathematically defined as $C = A \times B$, where $A$ is a $M \times K$ sparse matrix, $B$ is a $K \times N$ dense matrix, and $C$ is the resulting $M \times N$ dense matrix.

\textbf{Vectorized SpMM}
Vectorized SpMM can be implemented through various approaches\cite{vectorspmm1,vectorspmm2,vectorspmm3}. The row-wise approach, also known as Gustavson’s algorithm\cite{gustavson}, has been shown to be highly effective. It computes each output row by accumulating parallel products of $A[i,k]$ with $B[k,:]$, as follows:

\begin{equation}
C[i, :] = \sum_{k} A[i, k]B[k, :]
\label{eq1}
\end{equation}

The row-wise approach can be easily vectorized since each element of $A$ is multiplied with all the elements of a row in matrix $B$. This yields a vector of partial results, which naturally aligns with the AIV execution paradigm.

\begin{figure}[tb]
    \centering
    \includegraphics[width=1\linewidth]{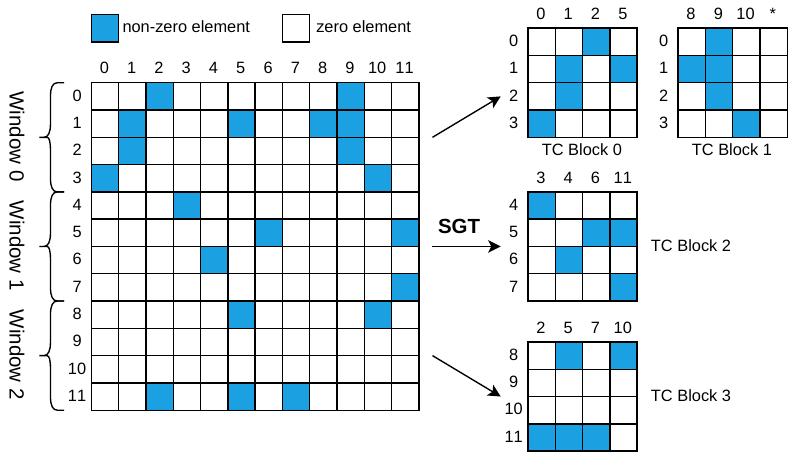}
    \caption{\small Sparse Graph Translation (SGT).}
    \label{fig2}
\Description{} \end{figure}

\textbf{Cube SpMM}
Methods designed for GPU Tensor Cores~\cite{tcu} are partially applicable to Cube Units. They require converting sparse matrices into dense TC blocks via Sparse Graph Translation (SGT) introduced by TC-GNN\cite{TCGNN}. As shown in Figure \ref{fig2}, we illustrate a simplified example with a block size of $4\times 4$. The nonzeros (blue squares) in the sparse matrix are partitioned into row windows. Within each window, every matrix column containing at least one nonzero is represented as a fixed-length column vector padded with zeros (white square). Each window may require additional zero-padding columns on the right side to form a complete TC block. In SpMM, these blocks serve as the left operand; the corresponding dense B blocks are loaded according to column indices of the A blocks. Results are accumulated on-chip and written back to global memory after completing a whole row window.

\subsection{Bottleneck and Opportunity}

To reveal the architectural bottlenecks of SpMM on Ascend NPUs, we examine the following two aspects.

\textbf{Write Conflicts in Heterogeneous Execution.}
Prior work has established that the physically decoupled AIC and AIV can execute concurrently and that coordinating their workloads is important for SpMM performance~\cite{neutronsparse}, yet the overhead caused by heterogeneous execution has not been considered: when the AIC and AIV concurrently write partial sums to the same rows of the output matrix $C$, atomic accumulation is required for correctness, and the resulting write conflicts can reduce performance.

\begin{figure}[t]
    \centering
    \includegraphics[width=\linewidth]{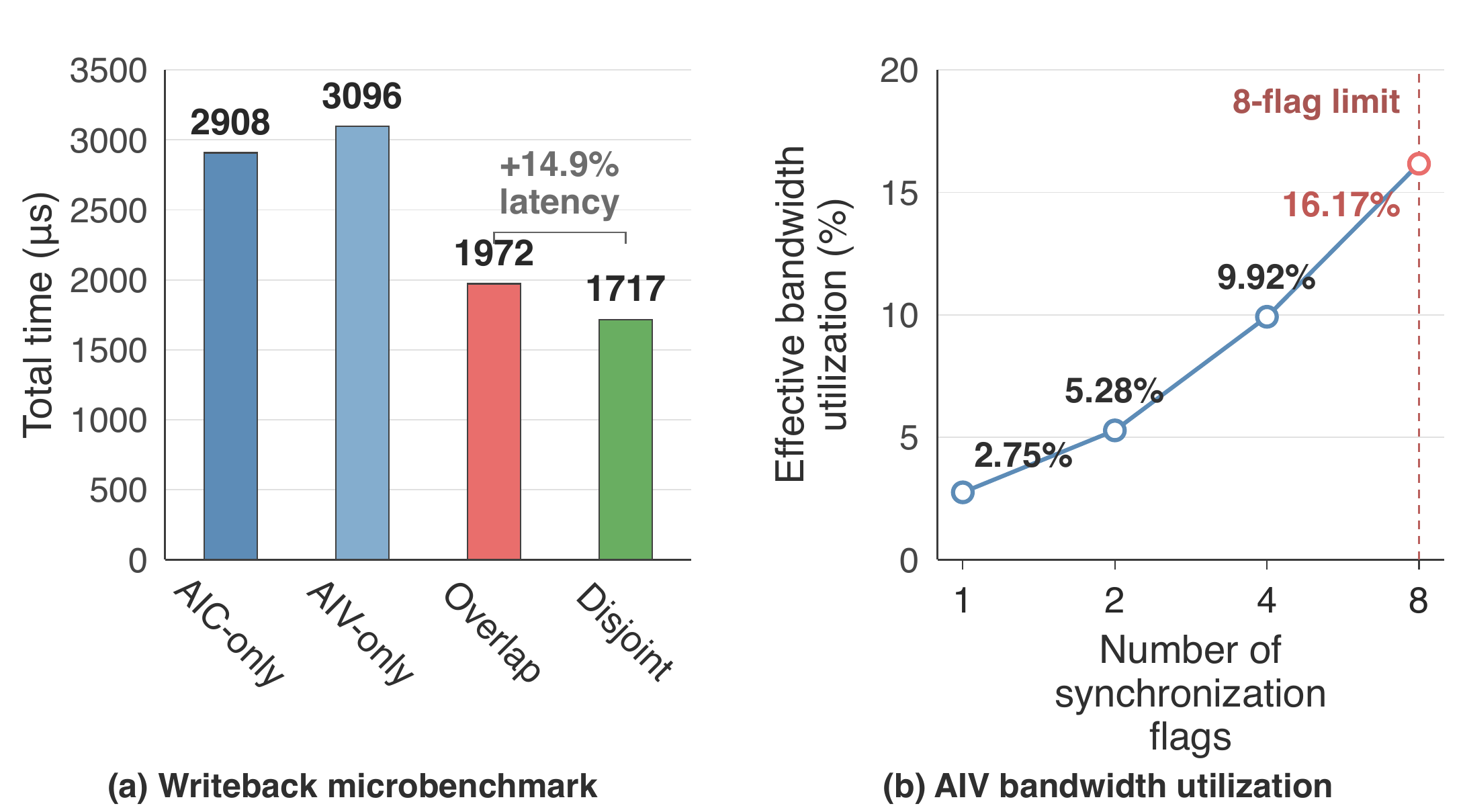}
    \caption{\small (a) Write-only microbenchmark. (b)Effective bandwidth utilization of the naive CSR SpMM with different numbers of synchronization flags on \textit{mip1}
    ($N=256$, FP32).}
    \label{fig:motivation}
\Description{} \end{figure}

To quantify this effect, we construct a write-only microbenchmark with four configurations: AIC-only, AIV-only, concurrent AIC-AIV execution with overlapping destinations, and concurrent AIC-AIV execution with disjoint destinations. Each run writes the same 640~KB of output data. In both concurrent configurations, the AIC and AIV each write half of the data; the configurations differ only in whether their destination regions overlap. We report the aggregate time of 1,000 consecutive runs. As shown in Figure~\ref{fig:motivation}(a), while concurrent execution reduces the total time compared to single-processor execution, overlapping destinations incur an additional 14.9\% overhead compared to disjoint destination. This result demonstrates the cost of atomic write conflicts and motivates limiting shared output ownership between the two types of processors. It suggests that heterogeneous workload mapping should consider not only load balance but also output ownership. By assigning each output row exclusively to one processor whenever possible and confining unavoidable AIC-AIV sharing to a bounded set, we can preserve the benefits of concurrent execution with less overhead.


\textbf{MTE Starvation Caused by Inefficient Software
Pipelining.}
The NPU relies on a single-threaded Scalar Unit to explicitly issue instructions and manage data movement across various computing units and multi-level memory hierarchy\cite{Ascend, davinci}. Because of this single-threaded execution model, an inefficient software pipeline inflicts severe MTE starvation, exposing different architectural bottlenecks on AIC and AIV.

\begin{table}[t]
  \centering
  \small
  \caption{\small AIC memory-hierarchy parameters on Ascend Atlas 800I A2 NPU (per AIC).}
  \label{tab:aic_memory_hierarchy}
  \small
  \setlength{\tabcolsep}{4pt}
  \begin{tabularx}{\linewidth}{@{}lXl@{}}
    \toprule
    \textbf{Parameter} & \textbf{Resource or Data Path}
                       & \textbf{Value} \\
    \midrule
    \multirow{3}{2cm}{Capacity}
      & L1 buffer                 & 512 KB \\
      & L0A/L0B operand buffers   & 64 KB each \\
      & L0C accumulation buffer   & 256 KB \\
    \midrule
    \multirow{4}{2cm}{Latency}
      & MTE2 (GM $\rightarrow$ L1)        & $>$ 100 ns \\
      & MTE1 (L1 $\rightarrow$ L0A/L0B)  & \textasciitilde 20 ns \\
      & Cube (Compute Unit)   & \textasciitilde 5 ns \\
      & Scalar (Dispatch Instructions)   & \textasciitilde 2 ns \\
    \bottomrule
  \end{tabularx}
\end{table}

In AIC, data movement and computation across the memory hierarchy exhibit substantially different latencies, while the associated buffers possess vastly different capacities, as summarized in Table~\ref{tab:aic_memory_hierarchy}. Without hierarchy-aware tiling and pipelining, transfer and computation stages become partially serialized: short-latency MTE1, Cube, and Scalar instructions delay the issuance of subsequent MTE2 requests, preventing MTE2 from maintaining enough in-flight transfers. As a result, MTE2 is frequently starved of requests, creating pipeline stalls and underutilizing its bandwidth. 

In AIV, the hardware imposes a strict limit of eight synchronization flags for coordinating the MTE and compute units. For the highly irregular AIV execution, eight in-flight memory requests are utterly insufficient to hide the long-tail memory latency, forcing the instruction stream into serialized execution. To quantify this limitation, we execute the naive CSR SpMM on \textit{mip1} with $N=256$ and FP32 while varying the number of synchronization flags, as shown in Figure~\ref{fig:motivation}(b). However, even at the hardware limit of eight flags, only 16.17\% of the effective bandwidth is utilized. This result shows that the limited number of synchronization flags bounds the number of in-flight memory requests and prevents the AIV from exposing sufficient memory-level parallelism.

The above architecture-specific bottlenecks demand NPU-friendly sub-kernel designs. By orchestrating explicit hierarchical pipeline on the AIC and introducing implicit synchronization technique on the AIV, we can effectively mask memory latency and substantially reduce pipeline bubbles.

\begin{figure}[tb]
    \centering
    \includegraphics[width=\linewidth]{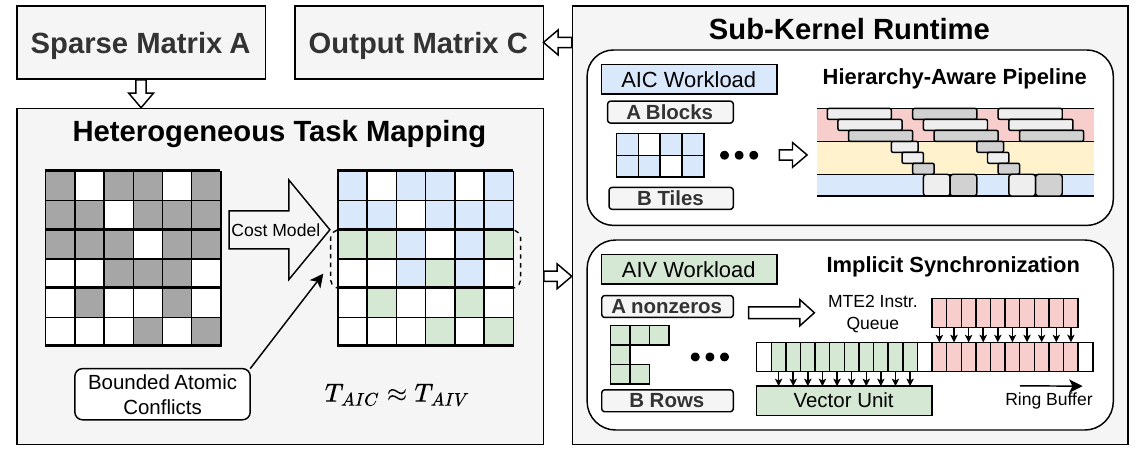}
    \caption{\small Overview of Ascend-SpMM.}
    \label{fig:overview}
\Description{} \end{figure}

\section{Ascend-SpMM}\label{sec3}

\subsection{Overview}

Ascend-SpMM is a novel systematic framework designed for sparse matrix-dense matrix multiplication, which maximizes the utilization of the heterogeneous resources on Ascend NPUs. Figure ~\ref{fig:overview} provides an overview of Ascend-SpMM. To achieve efficient hybrid SpMM computation on the decoupled NPU architecture, we first establish a memory-centric cost model to quantify the performance characteristics of the AI Cube (AIC) and AI Vector (AIV) when processing matrices with different density patterns. Based on the cost model, we propose a conflict-bounded density-aware heterogeneous task mapping strategy, which accurately partitions the sparse matrix into Cube-friendly regular blocks and Vector-friendly residuals with light coordination overhead. To execute the resulting partitioned workload efficiently, we further design two NPU-friendly sub-kernels: the AIC kernel exploits a hierarchy-aware pipeline with double buffering to overlap data movement and Cube computation, and the AIV kernel adopts an implicit synchronization based on a ring buffer, transforming irregular memory fetches of Gustavson’s algorithm\cite{gustavson} into a continuous, overlapped instruction stream that bypasses the hardware synchronization-flag limit.

\subsection{Memory-Centric Cost Model}\label{subsec31}

SpMM is a fundamentally memory-bound operator, and accessing the dense matrix $B$ dominates the data movement cost. We therefore model the AIC and AIV execution times using the amount of $B$ data they access. Other hardware-dependent effects, including cache behavior and instruction overhead, are absorbed into two constants, $\alpha$ and $\beta$.

We partition the sparse matrix $A$ into $m=\lceil M/W\rceil$ row windows, where $W=16$ matches the row dimension of the Cube unit. Let $b_{i,k}=A[iW:(i+1)W,k]$ denote a non-empty $W\times1$ block within $i$-th row window. Let $d_{i,k} \in [1,W]$ denote its number of nonzeros. The execution costs of
assigning this block to the AIC and AIV are modeled as
\begin{equation}
    T_{\mathrm{AIC}}(b_{i,j})=\alpha N,
    \qquad
    T_{\mathrm{AIV}}(b_{i,j})=\beta d_{i,j}N .
    \label{eq:block-cost}
\end{equation}
In our implementation, $\alpha$ and $\beta$ are calibrated once using the effective AIC bandwidth $B_{AIC}$ and AIV bandwidth $B_{AIV}$:
\begin{equation}
\alpha \propto \frac{1}{B_{\mathrm{AIC}}},
\qquad
\beta \propto \frac{1}{B_{\mathrm{AIV}}}.
\end{equation}

The AIC pays a fixed cost because the block is zero-padded,
whereas the AIV processes only its nonzeros. A dense block amortizes one AIC access to $B[k,:]$ over more nonzeros,
whereas a sparse block benefits from the AIV's fine-grained execution.
Consequently, dense blocks should be prioritized for the AIC and sparse
blocks for the AIV.

\subsection{Conflict-bounded Density-Aware Task Mapping}\label{subsec32}

At the finest granularity, each $W\times1$ block in $A$ could be ranked by its density, with dense blocks assigned to the AIC and sparse blocks assigned to the AIV according to our cost model. However, such a mapping may split many row windows to different types of processors. The AIC and AIV would then update the same rows of $C$, introducing atomic operation conflicts for every split window. To bound the atomic conflicts, we formulate task mapping at the row-window granularity.

Let window $i$ contain $c_i$ nonempty blocks and $z_i$ nonzeros,
and define its effective density as $\rho_i=z_i/c_i$. Under a
continuous relaxation, let $x_i\in[0,1]$ be the fraction of
window $i$ assigned to the AIC. Assuming that nonzeros are
uniformly distributed within each divisible window, the mapping
problem is
\begin{equation}
\begin{aligned}
    \min_{\mathbf{x}\in[0,1]^m}\;
    \max\Bigg\{
         \alpha N\sum_{i=1}^{m}x_i c_i, 
         \beta N\sum_{i=1}^{m}(1-x_i)z_i
    \Bigg\}.
\end{aligned}
\label{eq:mapping-objective}
\end{equation}
The two terms are the concurrent AIC and AIV execution costs,
respectively.

\begin{theorem}[Threshold Optimality]
\label{thm:threshold}
For the relaxation in Eq.~\eqref{eq:mapping-objective}, suppose
the windows are ordered such that
$\rho_1\geq\rho_2\geq\cdots\geq\rho_m$. There exists an optimal
solution with an index $k$ satisfying
\[
    x_i=1\;(i<k),\qquad
    x_k\in[0,1],\qquad
    x_i=0\;(i>k).
\]
Thus, at most one row window is shared by the AIC and AIV.
\end{theorem}

\begin{proof}
Consider two windows $i<j$ with $x_i<1$ and $x_j>0$.
Moving an equal AIC block budget $\epsilon$ from $j$ to $i$
preserves the AIC cost and changes the AIV cost by
$-\beta N\epsilon(\rho_i-\rho_j)\leq 0$. Repeating this exchange
produces a density-ordered prefix assignment. Consolidating any
fractional assignments with equal density leaves at most one
fractional window.
\end{proof}

\begin{figure}[tb]
    \centering
    \includegraphics[width=\linewidth]{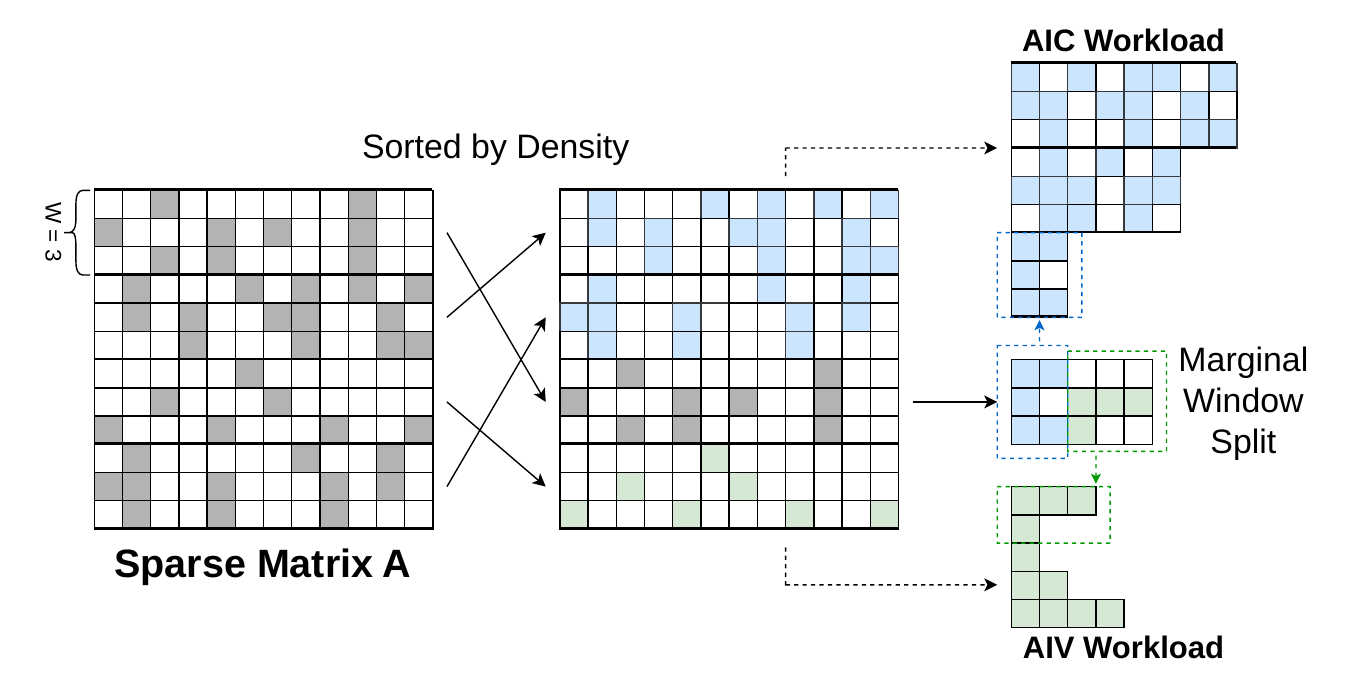}
    \caption{\small Cross-Core Task Mapping. Row windows are sorted by density, and only the marginal window is split to generate AIC and AIV workloads.}
    \label{fig:task_mapping}
\Description{} \end{figure}

After sorting the windows, a linear scan identifies the boundary
window $k$. We discretize only this window. Let its block
occupancies be sorted as
$d_{k,1}\geq\cdots\geq d_{k,c_k}$ and define
\[
    D_k(q)=\sum_{j=1}^{q}d_{k,j},\qquad
    C_k=\sum_{i<k}c_i,\qquad
    Z_k=\sum_{i>k}z_i .
\]
The number of boundary blocks assigned to the AIC is selected by
\begin{equation}
\begin{aligned}
q^\star=\arg\min_{0\leq q\leq c_k}\;&
N\max\Big\{
    \alpha(C_k+q),\,
    \beta\big[Z_k+z_k-D_k(q)\big]
\Big\},
\end{aligned}
\label{eq:boundary-refinement}
\end{equation}
Hence, cross-processor output sharing remains bounded
to at most one row window. Figure~\ref{fig:task_mapping} illustrates the task mapping strategy with $W=3$.

The selected AIC blocks are stored in $W\times1$ BSR format\cite{BSR},
whereas the AIV workload remains in CSR format. Window profiling
takes $\mathcal{O}(\mathrm{NNZ})$ time, sorting takes
$\mathcal{O}(m\log m)$, and the boundary window can be refined by
bucket sorting in $\mathcal{O}(c_k+W)$. The total preprocessing
complexity is therefore
$\mathcal{O}(\mathrm{NNZ}+m\log m)$, and the resulting mapping can
be reused across repeated SpMM executions.

\subsection{AIC Kernel with Hierarchy-Aware Pipelining}\label{subsec33}

\begin{figure}[tb]
    \centering
    \includegraphics[width=\linewidth]{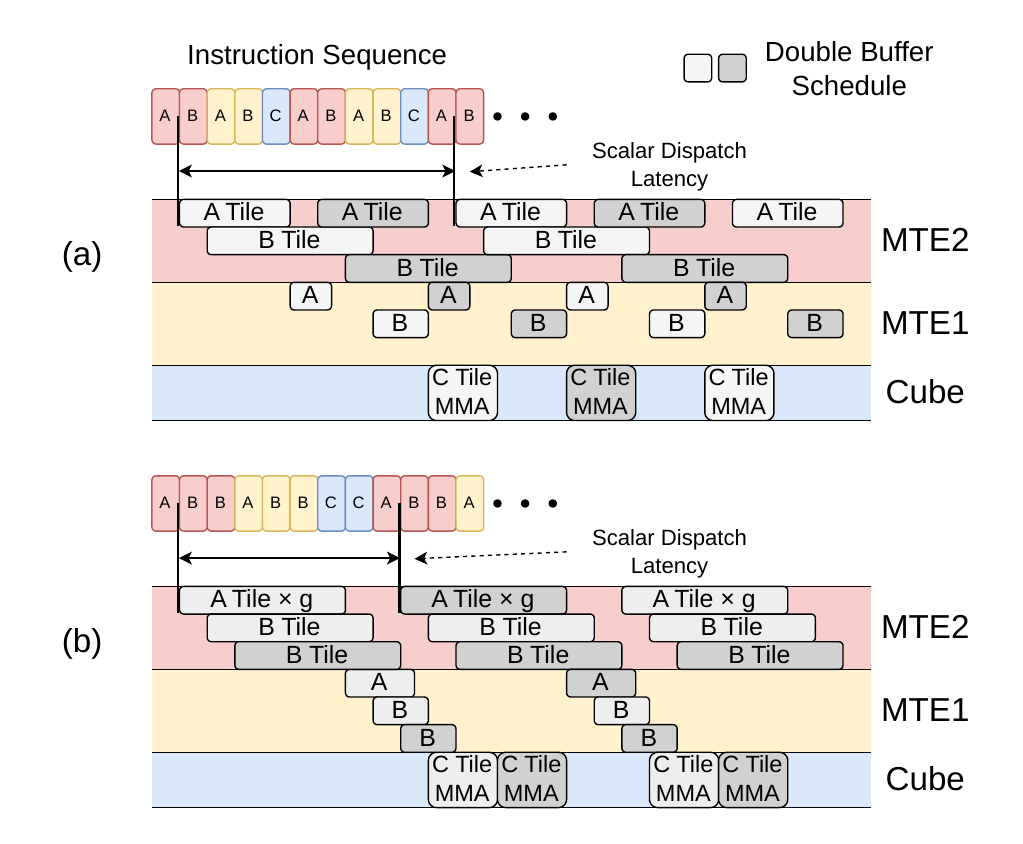}
    \caption{\small AIC pipeline schedules. (a) Naive tile-by-tile scheduling
interleaves MTE2 requests with MTE1 and Cube instructions, introducing more MTE2 issue latency. (b) The hierarchy-aware pipeline with $g=2$ batches sparse tiles
and issues two dense-tile transfers in advance, allowing MTE2
transfers to overlap with inner-level instruction dispatch and execution.}
    \label{fig3}
\Description{} \end{figure}

To improve utilization of the Matrix
Multiply-Add (Cube) Unit on the Ascend NPU, we design
a highly customized AIC sub-kernel tailored to the explicit multi-level memory hierarchy
(Global Memory $\rightarrow$ L1 $\rightarrow$ L0A/L0B $\rightarrow$
L0C). The kernel uses $W\times1$ BSR blocks as the basic storage in Global Memory. We first select a tile shape that
reduces padding overhead while respecting the on-chip buffer
constraints, then introduce a hierarchy-aware pipeline that
decouples the coarse-grained MTE2 transfers from the finer-grained
MTE1 and Cube operations.

\textbf{Hierarchy-Aware Tiling.}
Let $(m,k,n)$ denote the tile dimensions for multiplying an
$m\times k$ sparse tile with a $k\times n$ dense tile. As observed by
previous work~\cite{flashsparse}, reducing $m$ decreases the number of
padded zeros within a sparse tile and therefore reduces redundant Cube
computation. We consequently set $m=W=16$, which is the smallest Cube-compatible tile height and matches the
height of one row window.

For a fixed $k$, increasing $n$ reduces the number of tiles along the
feature dimension of $B$. This reduces the number of MTE2 transfer
instructions for $B$, improves per-instruction throughput by amortizing the fixed launch overhead.
We therefore maximize $n$ under the on-chip buffer constraints. Since
the minimum Cube-compatible reduction dimension is $k=16$, the
double-buffered L0B capacity imposes $2kns_B \leq S_{\mathrm{L0B}}$, 
where $s_B$ is the element size and
$S_{\mathrm{L0B}}=64$~KB, which indicates $n= 1024$ under FP16 and $n= 512$ under FP32.

\textbf{Hierarchy-Aware Pipeline.}
After fixing the tile size, we
decouple the scheduling granularities of MTE2 and the inner pipeline.
Instead of processing one tile at a time, MTE2 stages a batch of tiles in L1, while MTE1 and the Cube Unit continue to consume the
staged data tile by tile.

Figure~\ref{fig3}(a) illustrates a naive tile-by-tile pipeline. For
each tile, the Scalar Unit dispatches its MTE2, MTE1, and Cube
instructions to corresponding hardware units before proceeding to the next tile. Because instruction
dispatch is single-threaded, the MTE1 and Cube instructions inserted
between consecutive MTE2 transfers delay the issue of the next
GM-to-L1 requests. These issue latencies limit the number of in-flight MTE2
transfers and result in low MTE2 bandwidth utilization.

As shown in Figure~\ref{fig3}(b), we instead batch multiple tiles at the
MTE2 level. The AIC workload consists of $W\times 1$ blocks,
which can be concatenated into a larger $W\times(gk)$ tile
for a batch of $g$ reduction tiles. This reduces the number of
MTE2 instructions accessing $A$. The corresponding B-tile transfers are also
issued consecutively before dispatching their MTE1 and Cube
instructions. By issuing multiple MTE2 requests in
advance, their transfers overlap with the subsequent dispatch and
execution of MTE1 and Cube instructions, reducing MTE2 issue latency and
improving its effective bandwidth utilization. Once the data is staged in L1, MTE1 and the Cube Unit
consume it at the original $k=16$ computation granularity. We further restrict $g$ to be even to accommodate double buffering. Taking into account the constraint of L1=512KB, we choose $g=6$ in practice.

\subsection{AIV Kernel with Implicit Synchronization}\label{subsec34}

\begin{algorithm}[tb]
\caption{\small AIV Round-Robin Schedule within a row.}\label{algo3}
\begin{algorithmic}[1]

\Require row\_ptr, col\_id, value, $B$, row\_id
\Ensure Result Matrix $C$

\State \texttt{\_\_ub\_\_} b[$L \times N$], c[$N$];

\State start = row\_ptr[row\_id], end = row\_ptr[row\_id + 1];

\State \text{duplicate}(c, 0, $N$);
\For{t = start \textbf{ to } end}
    \State slot = t mod L;
    \State \text{copy\_gm\_to\_ub}(b[slot $\times N$], $B$[col\_id[t] $\times N$] , $N$);
    
    \State \text{set\_flag}(\texttt{PIPE\_MTE2}, \texttt{PIPE\_V}, 0);
    \State \text{wait\_flag}(\texttt{PIPE\_MTE2}, \texttt{PIPE\_V}, 0);
    \State \text{vaxpy}(c, b[slot $\times N$], value[t], $N$);
\EndFor
\State \text{copy\_ub\_to\_gm}($C$[row\_id $\times N$], c, $N$);
\end{algorithmic}
\end{algorithm}

\begin{figure}[tb]
    \centering
    \includegraphics[width=\linewidth]{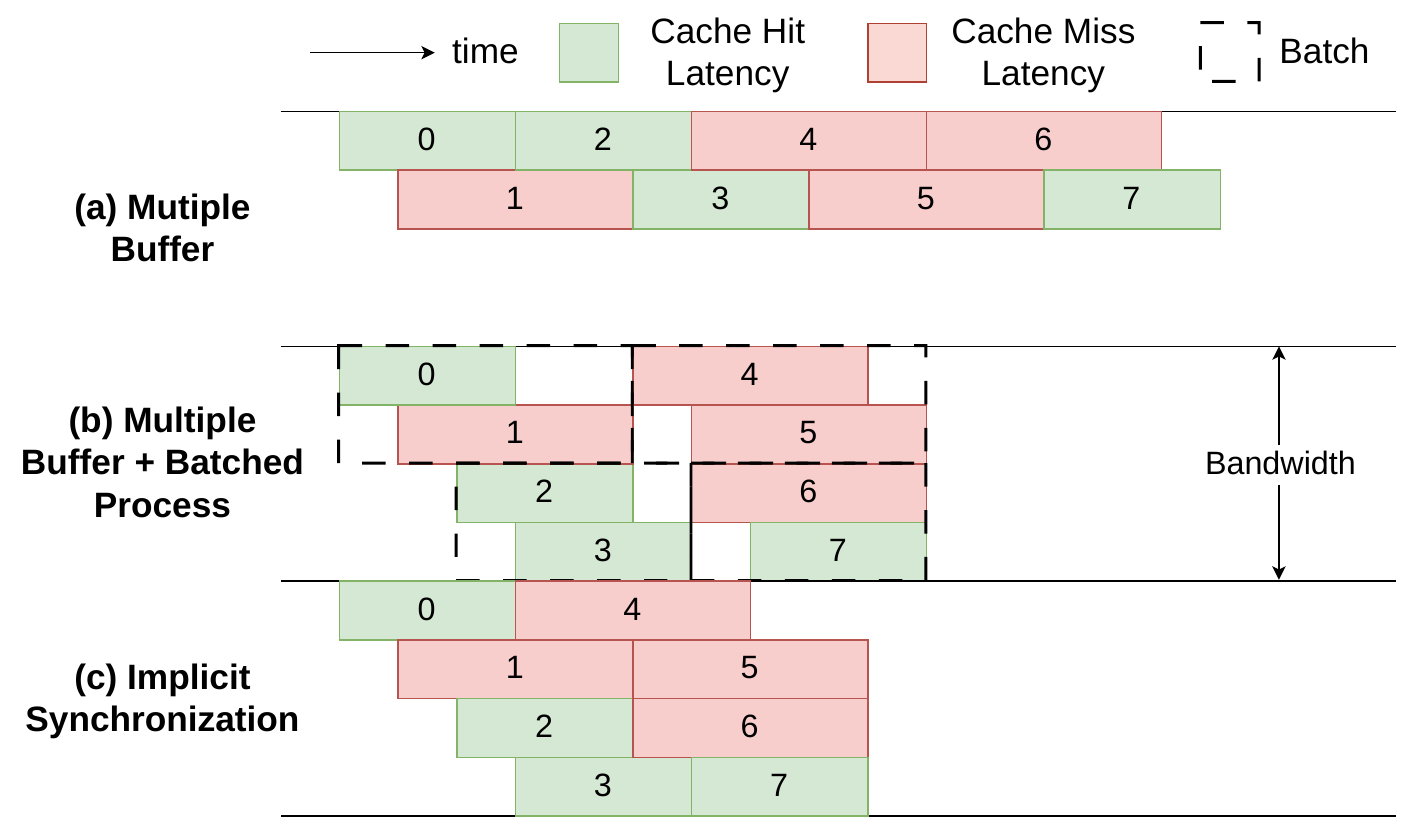}
    \caption{\small Simplified MTE2 bandwidth utilization timeline processing eight row fetches from the dense matrix. (a) Traditional Multiple Buffer strategy with buffer number = 2. (b) Multiple Buffer + Batched Process strategy with buffer num = 2 and batch size = 2. (c) Implicit Synchronization strategy.}
    \label{fig:AIV}
\Description{} \end{figure}

The AIV kernel processes irregular long-tail rows directly in the CSR
format. Following Gustavson's algorithm~\cite{gustavson}, each nonzero element triggers a
fine-grained row fetch $D_t$ from the dense matrix $B$, followed by a Vector
\texttt{vaxpy} operation $V_t$. Such short and random transfers expose insufficient
memory-level parallelism to fully utilize MTE2.

\textbf{Synchronization Bottleneck.}
A conventional multi-buffer schedule protects each buffer in UB with both a
MTE2-to-Vector forward dependency, which ensures that computation starts
after data arrival, and a Vector-to-MTE2 backward dependency, which prevents
the next MTE2 instruction from overwriting the current computation operands. Since the hardware provides only eight synchronization flags, explicit synchronization restricts the number of memory transactions that can be processed simultaneously. Batching can increase the number of issued transfers, but a
batch cannot progress until its slowest transfer completes. Consequently,
the latency variation between L2 hits and misses introduces tail-induced
MTE2 bubbles, as illustrated in Figure~\ref{fig:AIV}(a) and (b).

\begin{figure*}[tb]
    \centering
    \includegraphics[width=\textwidth]{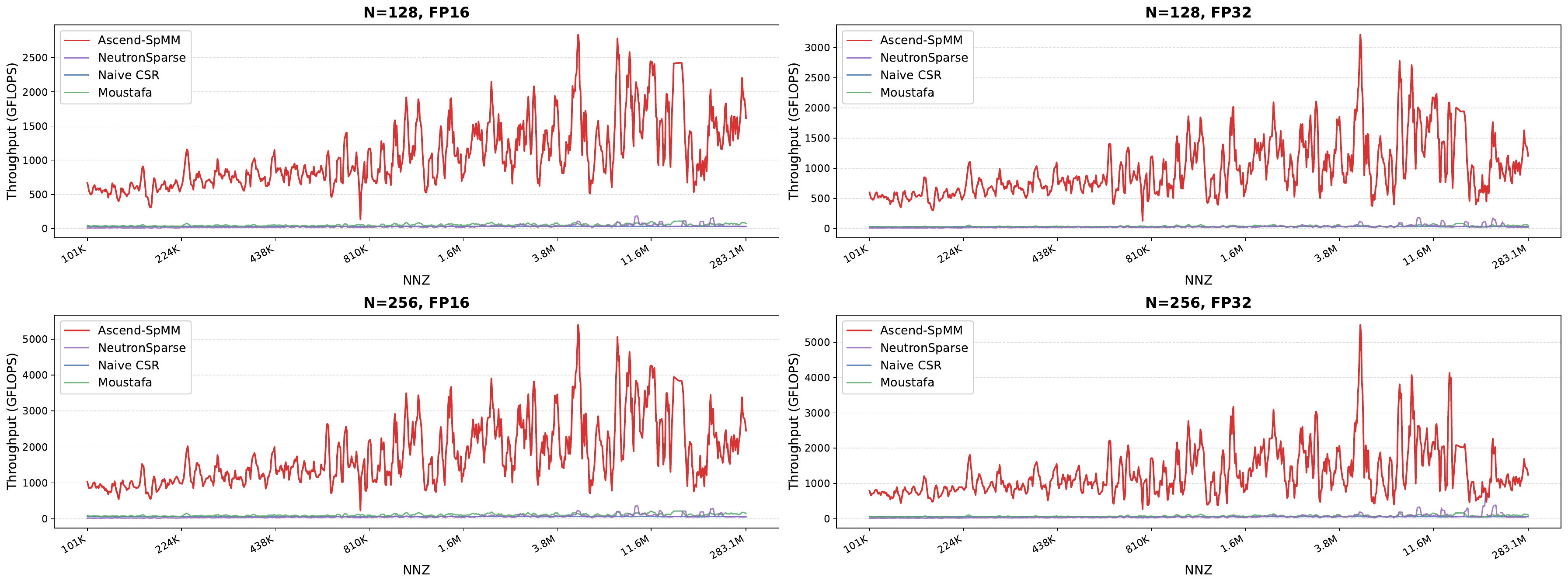}
    \caption{\small \textbf{SpMM performance results on NPU.} The matrices are sorted in ascending order by NNZ and each point in the figure represents the average GFLOPS for a contiguous set of 5 matrices. NeutronSparse has some missing data points due to OOM.}
    \label{fig:gflops_npu}
\Description{} \end{figure*}

\textbf{Implicit Backward Synchronization.}
The fundamental bottleneck of the conventional schedule is not the UB
capacity, but the limited number of synchronization flags. Each explicitly
managed buffer requires forward synchronization to ensure data readiness and
backward synchronization to protect its reuse, restricting the number of
in-flight buffers to the eight available flags. We therefore manage data transfer and computation by implicit backward synchronization, removing the explicit per-buffer Vector-to-MTE2 backward synchronization and instead
guaranteeing safe reuse through the maximum depth of the Vector instruction queue.

Let $Q_V$ denote the maximum depth of the Vector instruction queue. We use a ring buffer of $L$ buffers for row fetch and vector operation. When the Vector instruction queue is full, the Scalar unit will stop issuing subsequent instructions, until the first instruction to enter the queue is completed~\cite{Ascend}. The hardware
also ensures that $V_t$ is completed before $V_{t+1}$ completes, therefore $V_t$ is completed before $V_{t+Q_{V}}$ issued. Since $D_{t+Q_{V}+1}$ is issued after $V_{t+Q_{V}}$, choosing $L\ge Q_{V}+1$
ensures that the previous Vector instruction is completed before it is reused by MTE2, satisfying the backward synchronization requirement. Since instructions complete in program order, the remaining forward
dependencies can sequentially reuse a single flag~\cite{davinci}. Consequently, one flag
can coordinate a $L$-slot ring buffer with safe reuse, decoupling the prefetch distance from
the number of hardware synchronization flags.

As shown in Algorithm~\ref{algo3}, we organize the UB as a ring buffer with
$L$ slots (Line 1) and map the $t$-th dense-row fetch to slot
$s(t)=t\bmod L$ (Line 5). MTE2 continuously issues transfers to successive slots,
while the Vector pipeline performs the corresponding \texttt{vaxpy} operations after
the forward dependency is satisfied (Line 6-9). By removing batch-wide barriers and
backward dependencies, our design converts fine-grained random fetches into
a dense MTE2 instruction stream, reducing bandwidth bubbles and improving
sustained MTE2 utilization, as shown in Figure~\ref{fig:AIV}(c).

\section{Evaluation}

\subsection{Experimental Setup}

\textbf{Baselines.} We compare Ascend-SpMM with state-of-the-art methods on both NPUs and GPUs. For SpMM on Ascend NPUs, we compare against (i) an AIV-based SpMM implementation of Gustavson's algorithm\cite{gustavson}, denoted as Naive CSR; (ii) Moustafa\cite{moustafa}, an AIC-based SpMM method; and (iii) NeutronSparse\cite{neutronsparse}, an AIC-AIV collaborative SpMM method. Since no ready-to-use SpMM implementations for naive CSR and Moustafa are available on Ascend NPUs, we implemented both baselines following their published descriptions and tuned them on the target platform. For SpMM on GPUs, we compare with (i) CUDA Core-based methods, cuSPARSE\cite{cusparse} and RoDe\cite{RoDe}; (ii) Tensor Core-based methods, DTC-SpMM\cite{DTCSpMM} and FlashSparse\cite{flashsparse}; (iii) hybrid CUDA-core/Tensor-core method, HC-SpMM\cite{hcspmm}. For end-to-end GNN frameworks, we only compare with PyG\cite{pyg} and MindSpore-GL\cite{mindgl}, as DGL\cite{DGL} is not supported on Ascend NPUs. We evaluate the latest open-source implementations of these methods as strong baselines. The kernels are evaluated in single (FP32) and half precision (FP16), with the width of the dense matrices set to 128 and 256 (N = 128, 256).

\begin{table}[b]
\centering
\small
\caption{\small Detailed hardware specifications.}
\label{table4}
\begin{tabularx}{\linewidth}{@{}lXX@{}}
\toprule
\textbf{Specification} & \textbf{Ascend Atlas 800I A2} & \textbf{NVIDIA A100-SXM-40GB} \\
\midrule
Architecture          & Da Vinci               & Ampere                        \\
Compute Units         & 60 Cores               & 108 SMs                       \\
                      & (20 AIC + 40 AIV)      & (6912 CUDA Cores + 432 Tensor Cores) \\
Memory Capacity       & 32 GB                  & 40 GB                               \\
Memory Bandwidth      & 800 GB/s               & 1555 GB/s                     \\
Peak Performance      & 280 TFLOPS @ FP16      & 312 TFLOPS @ FP16             \\
\bottomrule
\end{tabularx}
\end{table}

\textbf{Datasets.} We select 820 square matrices from the SuiteSparse~\cite{suitesparse} collection with at least 10K rows, 10K columns, and 100K nonzeros. These matrices cover a variety of sparsity patterns and are used to evaluate kernel performance. In addition, we select classic graph datasets Flickr\cite{yelp}, Yelp\cite{yelp}, and ogbn-arxiv\cite{ogb} for end-to-end GNN performance evaluation.

\textbf{Environment.} We conducted our experiments on an Ascend Atlas 800I A2 with CANN 8.5.0~\cite{CANN} for NPU experiments, and an NVIDIA A100-SXM-40GB GPU~\cite{Ampere} with CUDA 12.2 for GPU experiments. Detailed hardware specifications~\cite{A100spec, 910B4spec} are listed in Table~\ref{table4}. All tests were run ten consecutive times with warm-ups to obtain the average values using \texttt{msprof}~\cite{msprof}.

\textbf{Parameter Selection.} We select the parameters on the target platform: the measured AIC/AIV bandwidths are 295/505 GB/s, yielding $\alpha=1/295$ and $\beta=1/505$. Since the Vector queue depth is undocumented, we characterize it using the official \texttt{msprof}~\cite{msprof} as $Q_V=32$ and conservatively set $L=64>Q_V$.

\begin{table}[tb]
  \centering
  \small
  \caption{\small Speedup distribution of Ascend-SpMM over the strongest NPU baseline for each matrix.}
  \label{tab:npu_speedup_distribution}
  \begin{tabularx}{\linewidth}{l|XXXX}
    \toprule
    \textbf{Speedup} & \textbf{128 FP16} & \textbf{128 FP32} & \textbf{256 FP16} & \textbf{256 FP32} \\
    \midrule
    $<1\times$ (\%)
      & 0.0\% & 0.0\% & 0.0\% & 0.0\% \\
    $[1,3)$ (\%)
      & 0.4\% & 0.1\% & 0.5\% & 0.2\% \\
    $[3,10)$ (\%)
      & 4.9\% & 6.1\% & 17.8\% & 30.6\% \\
    $\geq 10\times$ (\%)
      & 94.8\% & 93.8\% & 81.7\% & 69.1\% \\
    \textbf{Geo. Mean} ($\times$)
      & \textbf{19.03}$\times$
      & \textbf{20.55}$\times$
      & \textbf{14.98}$\times$
      & \textbf{12.70}$\times$ \\
    \bottomrule
  \end{tabularx}
\end{table}

\begin{table}[tb]
  \centering
  \small
  \caption{\small SpMM Profile Analysis on Matrix \textit{mip1} ($66{,}463 \times 66{,}463$, NNZ$=10{,}352{,}819$), $N=256$, FP32}
  \label{tab:mip1_profile}
  \begin{tabularx}{\linewidth}{l|XXX}
    \toprule
    Metric & Naive CSR & Moustafa & \textbf{Ascend-SpMM} \\
    \midrule
    Exec.\ Time ($\mu$s)   & 95854.44 & 58795.06 & 1350.50 \\
    AIC MTE2 Ratio (\%)    & ---      & 24.0\%   & 94.6\%   \\
    AIV MTE2 Ratio (\%)    & 34.8\%   & ---      & 93.4\%   \\
    \bottomrule
  \end{tabularx}
\end{table}

\subsection{SpMM Performance Comparison on Ascend NPU}

Figure~\ref{fig:gflops_npu} compares the throughput of Ascend-SpMM with the NPU
baselines across 820 matrices sorted by NNZ. Ascend-SpMM maintains
higher throughput over the full NNZ range. For a conservative comparison, Table~\ref{tab:npu_speedup_distribution}
reports the speedup over the strongest NPU baseline selected separately
for each matrix and configuration. Ascend-SpMM outperforms this
per-matrix baseline on every evaluated matrix, achieving geometric-mean
speedups of $19.03\times$, $20.55\times$, $14.98\times$, and
$12.70\times$ under 128 FP16, 128 FP32, 256 FP16, and 256 FP32,
respectively. Across the four configurations, 99.5--99.9\% of the
matrices achieve at least a $3\times$ speedup, while 69.1--94.8\%
exceed $10\times$; only 0.1--0.5\% fall between $1\times$ and
$3\times$. Although the relative gain decreases as $N$ increases,
Ascend-SpMM still delivers a $12.70\times$ geometric-mean speedup in
the most conservative 256 FP32 configuration, demonstrating robust
performance across matrix scales, feature widths, and data types.

To explain the above speedup, we profile the kernels on the matrix ``\textit{mip1}'', which contains both long and short rows and is representative of the sparse patterns of the majority of matrices. We report execution time and the AIC/AIV MTE2 utilization ratios. As shown in Table~\ref{tab:mip1_profile}, naive CSR executes exclusively on the AIV (Vector) cores with an AIV MTE2 ratio of merely 34.8\%, while Moustafa runs solely on the AIC (Cube) cores with an AIC MTE2 ratio of only 24.0\%, indicating that both baselines leave the MTE severely under-utilized. In contrast, our method employs a carefully designed software pipeline that overlaps computation and data movement within both the AIC and AIV pipelines. This overlap raises the AIC MTE2 ratio to 94.6\% and the AIV MTE2 ratio to 93.4\%, which in turn delivers an order-of-magnitude increase in effective memory bandwidth. Meanwhile, the cooperative use of both Cube and Vector units fully exploits the heterogeneous hardware resources. The combined effect explains the speedup observed over the baselines.

\begin{figure}[tb]
    \centering
    \includegraphics[width=\linewidth]{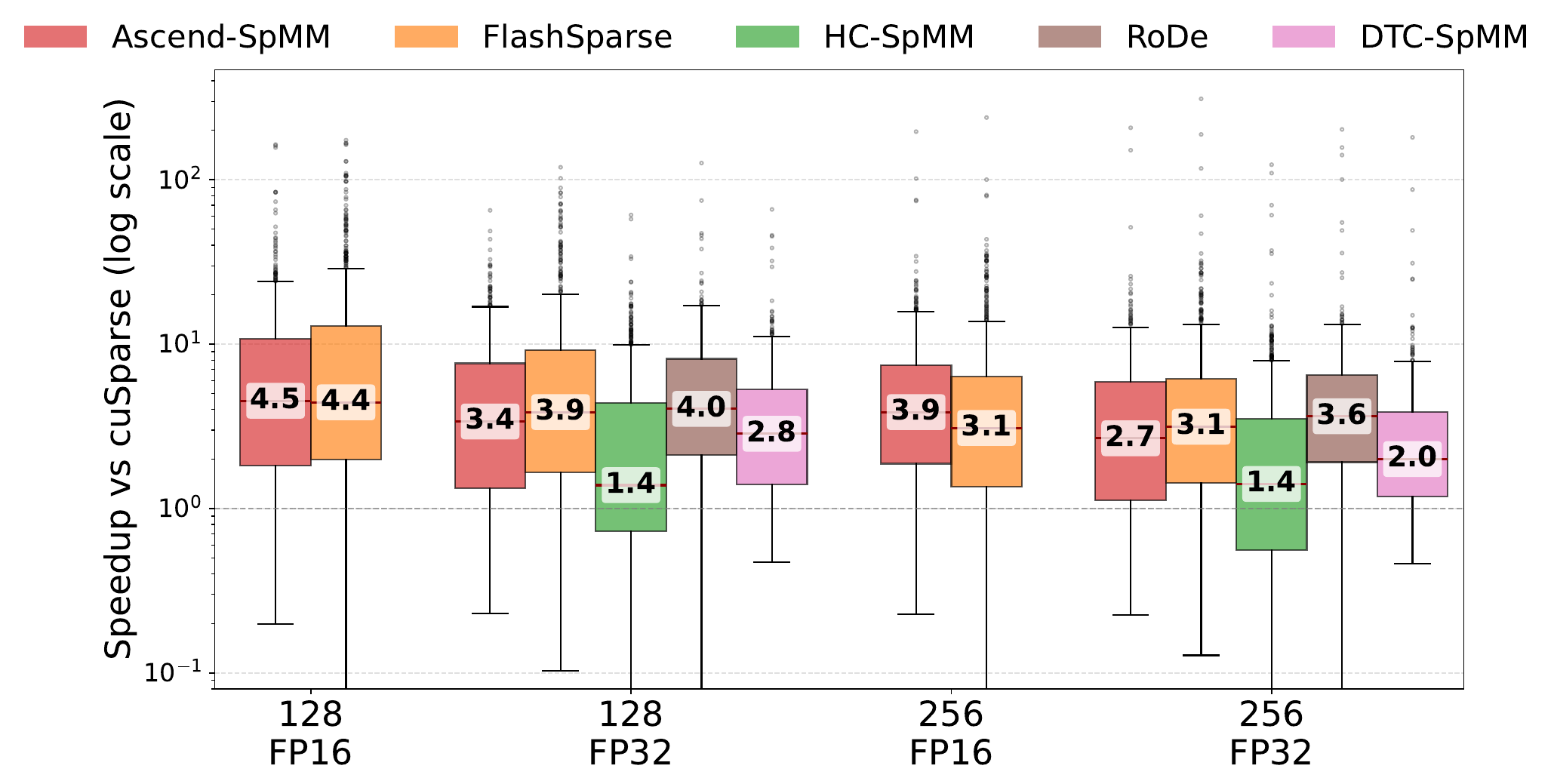}
    \caption{\small The speedup distribution of Ascend-SpMM and baselines over cuSPARSE. The number in the box plot represents the median speedup.}
    \label{fig:gpu_speedup}
\Description{} \end{figure}

\begin{table*}[tb]
  \centering
  \small
  \caption{\small Detailed Speedup Distribution vs cuSPARSE. Some GPU methods do not support FP16 computation.}
  \label{tab:gpu_speedup}
  \begin{tabularx}{\textwidth}{l|XX|XXXXX|XX|XXXXX}
    \toprule
    \textbf{Metric} & \multicolumn{2}{c|}{\textbf{N=128 FP16}} & \multicolumn{5}{c|}{\textbf{N=128 FP32}} & \multicolumn{2}{c|}{\textbf{N=256 FP16}} & \multicolumn{5}{c}{\textbf{N=256 FP32}} \\
      & Ascend-SpMM  & Flash Sparse  & Ascend-SpMM  & Flash Sparse  & HC-SpMM  & RoDe  & DTC-SpMM  & Ascend-SpMM  & Flash Sparse  & Ascend-SpMM  & Flash Sparse  & HC-SpMM  & RoDe  & DTC-SpMM \\
    \midrule
    $<1\times$ (\%)
      & 9.9\% & 11.0\%
      & 18.1\% & 11.7\% & 39.7\% & 6.1\% & 13.8\%
      & 10.7\% & 15.0\%
      & 22.2\% & 11.8\% & 41.5\% & 8.4\% & 19.8\% \\
    $[1,1.5)$ (\%)
      & 8.2\% & 7.9\%
      & 11.7\% & 11.9\% & 12.9\% & 5.9\% & 14.4\%
      & 9.5\% & 12.8\%
      & 14.6\% & 16.0\% & 11.6\% & 4.6\% & 14.3\% \\
    $[1.5,2)$ (\%)
      & 8.5\% & 6.2\%
      & 7.4\% & 8.3\% & 7.1\% & 11.0\% & 9.4\%
      & 8.6\% & 9.4\%
      & 6.2\% & 9.0\% & 7.8\% & 15.5\% & 16.4\% \\
    $\geq 2\times$ (\%)
      & 73.4\% & 74.9\%
      & 62.7\% & 68.1\% & 40.3\% & 77.1\% & 62.3\%
      & 71.1\% & 62.9\%
      & 57.0\% & 63.2\% & 39.1\% & 71.5\% & 49.4\% \\
    \textbf{Geo. Mean} ($\times$)
      & 4.46$\times$ & \textbf{4.82}$\times$
      & 3.08$\times$ & \textbf{3.79}$\times$
      & 1.37$\times$ & 3.63$\times$ & 2.76$\times$
      & \textbf{3.61}$\times$ & 2.96$\times$
      & 2.48$\times$ & 2.94$\times$
      & 1.13$\times$ & \textbf{3.03}$\times$ & 2.12$\times$ \\
    \textbf{Min} ($\times$)
      & \textbf{0.199}$\times$ & 0.027$\times$
      & 0.181$\times$ & 0.035$\times$
      & 0.004$\times$ & 0.016$\times$ & \textbf{0.472}$\times$
      & \textbf{0.228}$\times$ & 0.046$\times$
      & 0.225$\times$ & 0.058$\times$
      & 0.007$\times$ & 0.020$\times$ & \textbf{0.461}$\times$ \\
    \textbf{Max} ($\times$)
      & 163.8$\times$ & \textbf{174.1}$\times$
      & 65.2$\times$ & 119.2$\times$
      & 61.1$\times$ & \textbf{126.4}$\times$ & 66.0$\times$
      & 196.2$\times$ & \textbf{239.0}$\times$
      & 207.2$\times$ & \textbf{310.6}$\times$
      & 123.5$\times$ & 202.4$\times$ & 181.0$\times$ \\
    \bottomrule
  \end{tabularx}
\end{table*}

\subsection{SpMM Performance Comparison with GPU Methods}

We further compare our NPU-based method against state-of-the-art GPU implementations, including FlashSparse, RoDe, DTC-SpMM, HC-SpMM, and cuSPARSE, on the same 820 square-matrix subset. Figure~\ref{fig:gpu_speedup} presents the speedup distribution. As shown in Table~\ref{tab:gpu_speedup}, our method achieves a geometric-mean speedup of 2.48$\times$--4.46$\times$ over cuSPARSE across the four configurations, with 57.0$\%$--73.4$\%$ of datasets exceeding 2$\times$ speedup. FlashSparse stands out as the strongest GPU baseline with a geometric-mean of 2.94$\times$--4.82$\times$ and 62.9$\%$--74.9$\%$ of datasets above 2$\times$, benefiting from higher raw GPU memory bandwidth and warp-level parallelism on regular sparse patterns. Our method surpasses FlashSparse in the 256 FP16 configuration (3.61$\times$ vs.\ 2.96$\times$), where the larger computational intensity better saturates the NPU's cube units and amortizes the software pipeline overhead.

Compared to the other GPU baselines, our method consistently outperforms DTC-SpMM and HC-SpMM, and achieves comparable results to RoDe. Notably, our method exhibits lower performance variance than FlashSparse: the minimum speedup of 0.199$\times$ (128 FP16) versus FlashSparse's 0.027$\times$, and only 9.9$\%$ of cases fall below 1$\times$ compared to 11.0$\%$ for FlashSparse, indicating our software pipeline provides more robust performance across diverse sparse patterns. These results demonstrate that a carefully designed NPU-based SpMM implementation can achieve competitive and, on certain workload profiles, higher throughput than several state-of-the-art GPU SpMM methods.

\begin{figure}[tb]
    \centering
    \includegraphics[width=\linewidth]{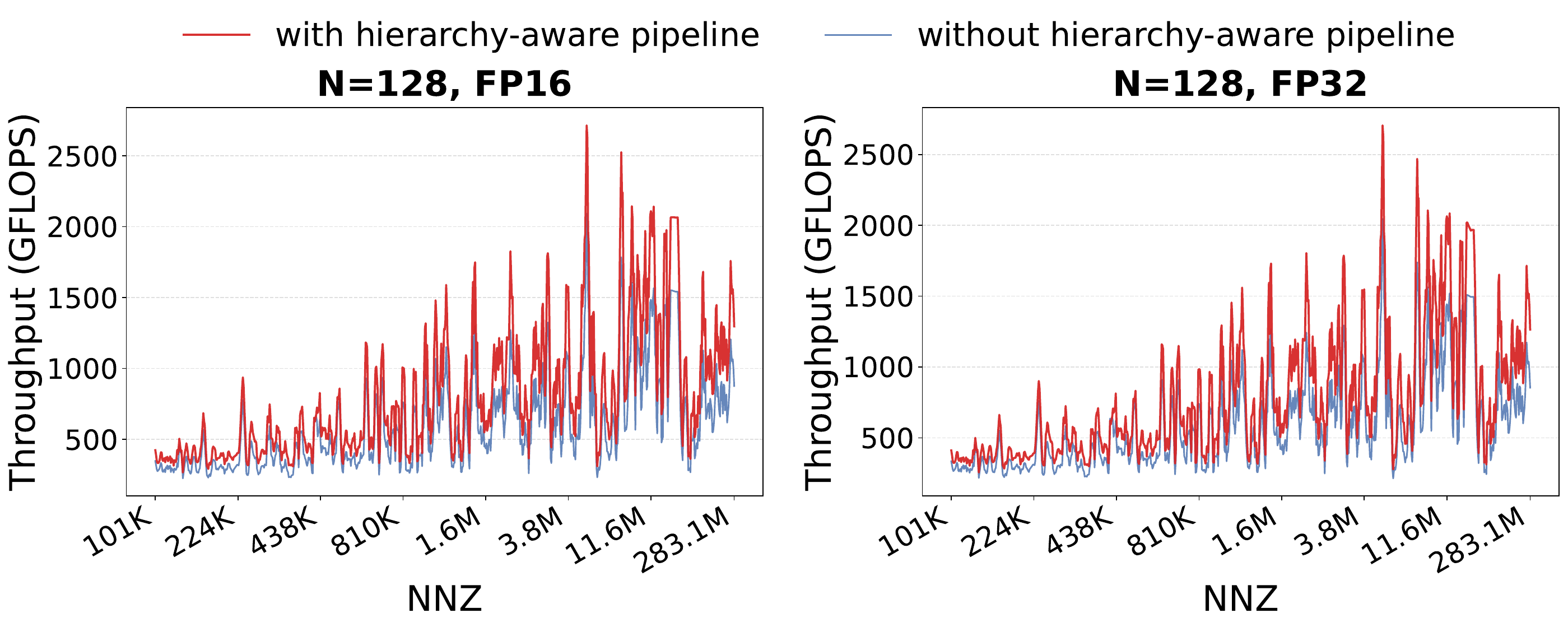}
    \caption{\small The throughput of AIC sub-kernel with/without Hierarchy-Aware Pipeline.}
    \label{fig:aic_pipeline}
\Description{} \end{figure}

\begin{figure}[tb]
    \centering
    \includegraphics[width=\linewidth]{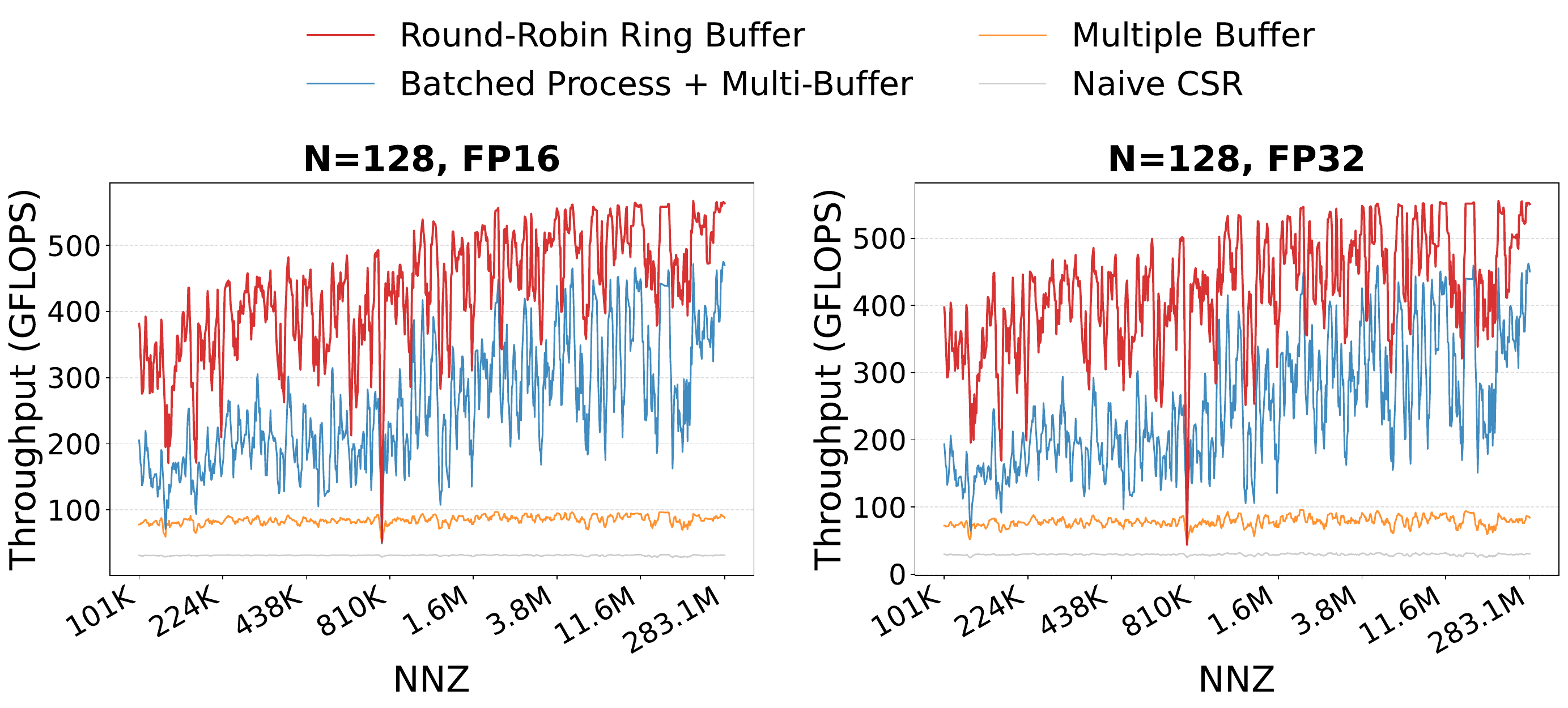}
    \caption{\small The throughput of AIV sub-kernel across various overlapping mechanisms.}
    \label{fig:aiv_pipeline}
\Description{} \end{figure}

\subsection{Ablation Study}

\textbf{Effectiveness of Hierarchy-Aware Pipeline.} We isolate the contribution of our hierarchy-aware pipeline on the AIC by comparing two AIC-only implementations: one with the optimized pipeline and one without. Figure ~\ref{fig:aic_pipeline} presents the GFLOPS comparison under N=128 FP16 and FP32 configurations on the 820 square-matrix subset. Enabling the hierarchy-aware pipeline achieves a geometric-mean speedup of 1.33× under FP16 and 1.31× under FP32. The uniform improvement across FP16 and FP32 further demonstrates that the pipeline's benefit is precision-agnostic and derives from reducing memory stalls rather than from compute-side optimizations.

\textbf{Effectiveness of Implicit Synchronization.} We evaluate the effectiveness of implicit synchronization on the AIV by comparing four AIV-only implementations: (i) Naive CSR, which processes sequentially with one buffer; (ii) Multiple Buffer with buffer number = 8, which decouples memory accesses from scalar execution at the per-element level; (iii) Batched Process + Multiple Buffer with batch size = 8 and buffer number = 8, which further amortizes loop overhead by grouping elements; and (iv) our AIV sub-kernel with $L = 64$ slots. Figure~\ref{fig:aiv_pipeline} shows the GFLOPS comparison under N=128 FP16 and FP32.

Compared with Naive CSR, our AIV sub-kernel achieves geometric-mean speedups of $12.87\times$ under FP16 and $13.43\times$ under FP32. Compared with Multiple Buffer, it remains $4.77\times$ and $5.11\times$ faster, respectively. Even over Batched Process + Multiple Buffer with the same total number of buffers, the round-robin ring buffer provides speedups of $1.81\times$ and $1.88\times$, reaching averages of 433 and 423 GFLOPS, respectively. The critical enabler of the speedup is implicit synchronization: the ring buffer allows each buffer slot to independently progress data transfer and computation without explicit barriers. This design eliminates inter-phase pipeline bubbles, which is essential on the AIV processor where the scalar-rich execution pattern of sparse row operations creates irregular compute latencies.

We also verified the correctness of implicit synchronization against the Naive CSR implementation with explicit forward and backward synchronization using a single buffer, and observed bitwise-identical results on all 820 matrices for $N\in \{16,32,64,128,256\}$ with random values in standard normal distribution $N(0,1)$.

\textbf{Effectiveness of Conflict-Bounded Density-Aware Task Mapping.} To quantify the contribution of conflict-bounded density-aware task mapping, we compare our strategy against three ablations: (i) a naive heterogeneous AIC+AIV hybrid that partitions rows by equally distributing the total NNZ count across cores; (ii) NeutronSparse\cite{neutronsparse} workload partition strategy; and (iii) Block-Level split that maps dense blocks to AIC and sparse blocks to AIV, ignoring atomic write conflicts. Table~\ref{tab:load_balance_ablation} summarizes the geometric-mean speedup of ours over each ablation across all three ablations.

\begin{table}[tb]
\centering
\small
\caption{\small Geometric-Mean Speedup of Ascend-SpMM over Heterogeneous Load Balancing Ablations.} \label{tab:load_balance_ablation} \begin{tabularx}{\linewidth}{l|XXXX}
\toprule
\textbf{Ours vs} & \textbf{128 FP16} & \textbf{128 FP32} & \textbf{256 FP16} & \textbf{256 FP32} \\
\midrule
Naive Load Balancing & 1.56× & 1.46× & 1.47× & 1.25× \\
NeutronSparse Load Balancing        & 1.36× & 1.29× & 1.26× & 1.23× \\
Block-Level Split & 1.19× & 1.20× & 1.14× & 1.10× \\
\bottomrule
\end{tabularx}
\end{table}

The proposed task mapping strategy consistently outperforms all ablations. Compared with naive NNZ-balanced mapping, it achieves geometric-mean speedups of $1.25\times$--$1.56\times$, as equal NNZ partitioning does not capture the distinct execution costs of AIC and AIV. Moreover, our strategy achieves a $1.23\times$--$1.36\times$ speedup over the NeutronSparse workload partitioning strategy, and a $1.10\times$--$1.20\times$ speedup over the Block-Level split strategy because these methods neglect the atomic write conflicts on output matrix. These results demonstrate that effective heterogeneous task mapping should consider both heterogeneous execution costs and write conflicts rather than workload volume alone.

\subsection{Case Study: End-to-end GNN Performance}

\begin{figure}[tb]
    \centering
    \includegraphics[width=\linewidth]{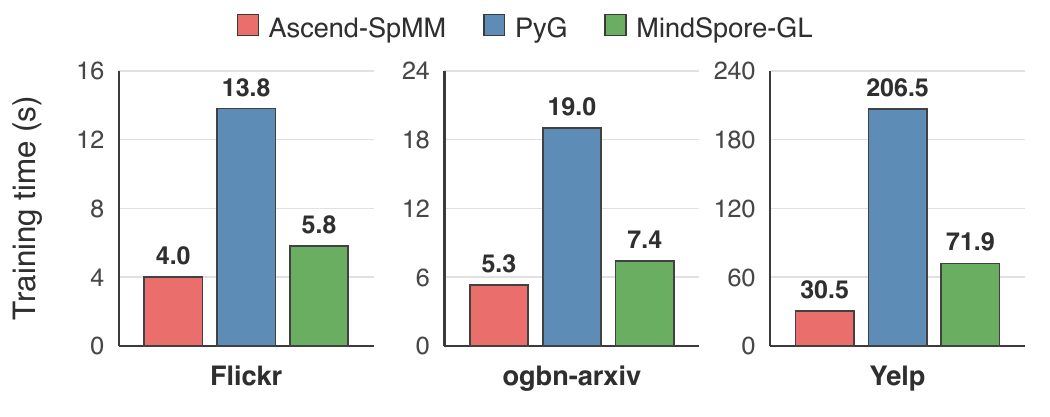}
    \caption{\small End-to-end two-layer GCN training time (200 epochs). Preprocessing time is included.}
    \label{fig:gcn_training}
\Description{} \end{figure}

To evaluate the benefits of our proposed Ascend-SpMM, we integrate it into PyG\cite{pyg} via the PyTorch C++ Extension\cite{paszke2019pytorch, torchnpu} interface. We compare our method against the original PyG\cite{pyg} implementation and MindSpore-GL\cite{mindgl} implementation on the Ascend Atlas 800I A2 NPU with the same two-layer Graph Convolutional Neural Network (GCN) model, replacing PyG’s message-passing operator with Ascend-SpMM. Due to the lack of native SpMM support on the NPU, both baselines resort to Scatter-based message passing rather than SpMM, which leads to substantial memory overhead and causes out-of-memory (OOM) failures on larger-scale graphs. 
We evaluate end-to-end model training time on three medium-scale datasets with a hidden dimension of 128. As shown in Figure~\ref{fig:gcn_training}, our method achieves speedups of 3.5$\times$--6.8$\times$ over PyG and 1.4$\times$--2.4$\times$ over MindSpore-GL across Flickr\cite{yelp}, ogbn-arxiv\cite{ogb}, and Yelp\cite{yelp}. The performance benefit widens considerably on Yelp---the largest graph of the three---where our approach attains a 6.8$\times$ speedup over PyG, demonstrating the scalability advantages of SpMM-based message passing over Scatter-based alternatives on NPUs.

\section{Related Work}

\textbf{SpMM Research on NVIDIA GPUs.} On traditional GPU CUDA cores, cuSPARSE\cite{cusparse} provides a widely used and high-performance baseline. Several CUDA-core
methods employ one-dimensional or adaptive tiling, row caching,
and row decomposition to improve memory access efficiency and load
balance~\cite{Sputnik, ASpT, gespmm, RoDe}. With hardware advancements, TC-GNN~\cite{TCGNN} pioneered the use of Tensor Cores~\cite{tcu} to accelerate sparse computations, and several subsequent studies explored optimization through reordering,
compact formats, and fine-grained packing~\cite{DTCSpMM, flashsparse, accspmm, generalsparse}. Recent studies have explored heterogeneous cooperation between CUDA cores and Tensor Cores~\cite{hcspmm, libra, brp}. However, these methods are designed for
the GPU execution model and do not address Ascend's independent
AIC/AIV memory paths, cross-core coordination, or scalar-issued
pipelines that require explicit synchronization.

\textbf{SpMM Research on Ascend NPUs.} Early NPU SpMM studies targeted coupled architectures. Moustafa et al.\cite{moustafa} explore multi-level tiling for
AIC-based SpMM, but irregular blocks still incur padding and the AIV
path remains unused in the modern decoupled architecture of Ascend NPUs. Recently, NeutronSparse\cite{neutronsparse} addressed this misalignment by introducing sparsity-aware workload coordination between heterogeneous engines and locality-aware tile orchestration. However, it does not optimize the kernel based on the underlying execution model of Ascend NPUs, which limits its overall performance.

\section{Conclusion}

This paper presents Ascend-SpMM, a systematic sparse matrix-dense matrix multiplication framework tailored to the decoupled architecture of Ascend NPUs. It combines a heterogeneous task mapping strategy with specialized AIC and AIV sub-kernels to exploit decoupled computing resources while reducing output conflicts and improving latency hiding. Extensive evaluations demonstrate that Ascend-SpMM achieves 12.70--20.55$\times$ geometric-mean speedup over the strongest evaluated NPU baseline, remains competitive with GPU implementations, and improves end-to-end GCN training performance. These results demonstrate the importance of jointly considering workload structure and architectural characteristics when optimizing sparse operators on heterogeneous accelerators.

\begin{acks}
In this paper, we use AI models: GPT 5.6, Gemini 3.1 Pro and Deepseek-V4-Pro, to polish writing and code debugging. All AI-generated or AI-revised content was rigorously reviewed, validated, and refined by the authors, who take full responsibility for the integrity, originality, and technical validity of the work presented.
\end{acks}

\bibliographystyle{ACM-Reference-Format}
\bibliography{refs}

@inproceedings{TCGNN,
  title={$\{$TC-GNN$\}$: Bridging sparse $\{$GNN$\}$ computation and dense tensor cores on $\{$GPUs$\}$},
  author={Wang, Yuke and Feng, Boyuan and Wang, Zheng and Huang, Guyue and Ding, Yufei},
  booktitle={2023 USENIX Annual Technical Conference (USENIX ATC 23)},
  pages={149--164},
  year={2023}
}

@inproceedings{Sputnik,
  title={Sparse gpu kernels for deep learning},
  author={Gale, Trevor and Zaharia, Matei and Young, Cliff and Elsen, Erich},
  booktitle={SC20: International Conference for High Performance Computing, Networking, Storage and Analysis},
  pages={1--14},
  year={2020},
  organization={IEEE}
}

@inproceedings{RoDe,
  title={A row decomposition-based approach for sparse matrix multiplication on GPUs},
  author={Pang, Meng and Fei, Xiang and Qu, Peng and Zhang, Youhui and Li, Zhaolin},
  booktitle={Proceedings of the 29th ACM SIGPLAN Annual Symposium on Principles and Practice of Parallel Programming},
  pages={377--389},
  year={2024}
}

@inproceedings{DTCSpMM,
  title={Dtc-spmm: Bridging the gap in accelerating general sparse matrix multiplication with tensor cores},
  author={Fan, Ruibo and Wang, Wei and Chu, Xiaowen},
  booktitle={Proceedings of the 29th ACM International Conference on Architectural Support for Programming Languages and Operating Systems, Volume 3},
  pages={253--267},
  year={2024}
}

@inproceedings{flashsparse,
  title={Flashsparse: Minimizing computation redundancy for fast sparse matrix multiplications on tensor cores},
  author={Shi, Jinliang and Li, Shigang and Xu, Youxuan and Fu, Rongtian and Wang, Xueying and Wu, Tong},
  booktitle={Proceedings of the 30th ACM SIGPLAN Annual Symposium on Principles and Practice of Parallel Programming},
  pages={312--325},
  year={2025}
}

@inproceedings{Ascend,
  title={Ascend: a scalable and unified architecture for ubiquitous deep neural network computing: Industry track paper},
  author={Liao, Heng and Tu, Jiajin and Xia, Jing and Liu, Hu and Zhou, Xiping and Yuan, Honghui and Hu, Yuxing},
  booktitle={2021 IEEE International Symposium on High-Performance Computer Architecture (HPCA)},
  pages={789--801},
  year={2021},
  organization={IEEE}
}

@inproceedings{davinci,
  title={Davinci: A scalable architecture for neural network computing},
  author={Liao, Heng and Tu, Jiajin and Xia, Jing and Zhou, Xiping},
  booktitle={2019 IEEE Hot Chips 31 Symposium (HCS)},
  pages={1--44},
  year={2019},
  organization={IEEE}
}

@manual{cusparse,
  author       = {NVIDIA Corporation},
  title        = {cuSPARSE Library User Guide},
  year         = {2024},
  url          = {https://docs.nvidia.com/cuda/cusparse/},
  note         = {Version 12.x}
}

@manual{tcu,
  author       = {NVIDIA Corporation},
  title        = {Tensor Core},
  year         = {2023},
  url          = {https://www.nvidia.com/en-us/data-center/tensor-cores/},
  note         = {Version 12.x}
}

@manual{Ampere,
  author       = {NVIDIA Corporation},
  title        = {NVIDIA Ampere GPU Architecture Tuning Guide},
  year         = {2023},
  url          = {https://docs.nvidia.com/cuda/ampere-tuning-guide/index.html},
  note         = {Version 12.x}
}

@book{CANN,
  title={Ascend AI Processor Architecture and Programming: Principles and Applications of CANN},
  author={Liang, Xiaoyao},
  year={2020},
  publisher={Elsevier}
}

@article{paszke2019pytorch,
  title={Pytorch: An imperative style, high-performance deep learning library},
  author={Paszke, Adam and Gross, Sam and Massa, Francisco and Lerer, Adam and Bradbury, James and Chanan, Gregory and Killeen, Trevor and Lin, Zeming and Gimelshein, Natalia and Antiga, Luca and Desmaison, Alban and Kopf, Andreas and Yang, Edward and DeVito, Zachary and Raison, Martin and Tejani, Alykhan and Chilamkurthy, Sasank and Steiner, Benoit and Fang, Lu and Bai, Junjie and Chintala, Soumith},
  journal={Advances in neural information processing systems},
  volume={32},
  year={2019}
}

@manual{torchnpu,
  author       = {Ascend},
  title        = {PyTorch Adaptation},
  year         = {2024},
  url          = {https://www.hiascend.com/en/developer/software/ai-frameworks/pytorch},
  note         = {Ascend Extension for PyTorch}
}

@manual{A100spec,
  author       = {NVIDIA Corporation},
  title        = {NVIDIA A100 | Tensor Core GPU},
  year         = {2023},
  url          = {https://www.nvidia.com/content/dam/en-zz/Solutions/Data-Center/a100/pdf/nvidia-a100-datasheet-us-nvidia-1758950-r4-web.pdf},
  note         = {A100 specification}
}

@manual{910B4spec,
  author       = {Ascend},
  title        = {Atlas 800I A2},
  year         = {2024},
  url          = {https://support.huawei.com/enterprise/zh/doc/EDOC1100360587/b2592910},
  note         = {Ascend 800I A2 specification}
}

@manual{msprof,
  author       = {Ascend},
  title        = {msProf},
  year         = {2024},
  url          = {https://www.hiascend.com/document/detail/zh/CANNCommunityEdition/850alpha001/devaids/optool/atlasopdev_16_0082.html}
}

@manual{BSR,
  author       = {NVIDIA Corporation},
  title        = {Block Sparse Row (BSR)},
  year         = {2023},
  url          = {https://docs.nvidia.com/cuda/cusparse/index.html#block-sparse-row-bsr}
}

@inproceedings{NPUarch1,
  title={Squeezing operator performance potential for the ascend architecture},
  author={Zhou, Yuhang and Wang, Zhibin and Liu, Guyue and Li, Shipeng and Lin, Xi and Wang, Zibo and Wang, Yongzhong and Wei, Fuchun and Zhang, Jingyi and Hu, Zhiheng and Liu, Yanlin and Li, Chunsheng and Zhang, Ziyang and Wang, Yaoyuan and Zhou, Bin and Dou, Wanchun and Chen, Guihai and Tian, Chen},
  booktitle={Proceedings of the 30th ACM International Conference on Architectural Support for Programming Languages and Operating Systems, Volume 2},
  pages={1156--1171},
  year={2025}
}

@inproceedings{NPUarch2,
  title={Accelerating Model Training on Ascend Chips: An Industrial System for Profiling, Analysis and Optimization},
  author={Zhou, Yuhang and Wang, Zibo and Wang, Zhibin and Zhang, Ruyi and Tian, Chen and Wang, Xiaoliang and Dou, Wanchun and Chen, Guihai and Wang, Bingqiang and Tian, Yonghong and Zhang, Yan and Wang, Hui and Wei, Fuchun and Sun, Boquan and Zhang, Jingyi and She, Bin and Su, Teng and Yao, Yifan and Li, Chunsheng and Zhang, Ziyang and Wang, Yaoyuan and Zhou, Bin and Liu, Guyue},
  booktitle={2025 USENIX Annual Technical Conference (USENIX ATC 25)},
  pages={1387--1408},
  year={2025}
}

@article{gustavson,
  title={Two fast algorithms for sparse matrices: Multiplication and permuted transposition},
  author={Gustavson, Fred G},
  journal={ACM Transactions on Mathematical Software (TOMS)},
  volume={4},
  number={3},
  pages={250--269},
  year={1978},
  publisher={ACM New York, NY, USA}
}

@article{vectorspmm1,
  title={Optimizing structured-sparse matrix multiplication in RISC-V vector processors},
  author={Titopoulos, Vasileios and Alexandridis, Kosmas and Peltekis, Christodoulos and Nicopoulos, Chrysostomos and Dimitrakopoulos, Giorgos},
  journal={IEEE Transactions on Computers},
  volume={74},
  number={4},
  pages={1446--1460},
  year={2025},
  publisher={IEEE}
}

@article{vectorspmm2,
  title={Sparse matrices in MATLAB: Design and implementation},
  author={Gilbert, John R and Moler, Cleve and Schreiber, Robert},
  journal={SIAM journal on matrix analysis and applications},
  volume={13},
  number={1},
  pages={333--356},
  year={1992},
  publisher={SIAM}
}

@inproceedings{vectorspmm3,
  title={Generalized sparse matrix-matrix multiplication for vector engines and graph applications},
  author={Li, Jiayu and Wang, Fugang and Araki, Takuya and Qiu, Judy},
  booktitle={2019 IEEE/ACM Workshop on Memory Centric High Performance Computing (MCHPC)},
  pages={33--42},
  year={2019},
  organization={IEEE}
}

@inproceedings{ASpT,
  title={Adaptive sparse tiling for sparse matrix multiplication},
  author={Hong, Changwan and Sukumaran-Rajam, Aravind and Nisa, Israt and Singh, Kunal and Sadayappan, Ponnuswamy},
  booktitle={Proceedings of the 24th Symposium on Principles and Practice of Parallel Programming},
  pages={300--314},
  year={2019}
}

@inproceedings{gespmm,
  title={Ge-spmm: General-purpose sparse matrix-matrix multiplication on gpus for graph neural networks},
  author={Huang, Guyue and Dai, Guohao and Wang, Yu and Yang, Huazhong},
  booktitle={SC20: International Conference for High Performance Computing, Networking, Storage and Analysis},
  pages={1--12},
  year={2020},
  organization={IEEE}
}

@inproceedings{accspmm,
  title={Acc-spmm: Accelerating general-purpose sparse matrix-matrix multiplication with gpu tensor cores},
  author={Zhao, Haisha and Li, San and Wang, Jiaheng and Zhou, Chunbao and Wang, Jue and Xin, Zhikuang and Li, Shunde and Liang, Zhiqiang and Pan, Zhijie and Liu, Fang and Zeng, Yan and Wang, Yangang and Chi, Xuebin},
  booktitle={Proceedings of the 30th ACM SIGPLAN Annual Symposium on Principles and Practice of Parallel Programming},
  pages={326--338},
  year={2025}
}

@inproceedings{generalsparse,
  title={$\{$GeneralSparse$\}$: Bridging the Gap in $\{$SpMM$\}$ for Pruned Large Language Model Inference on $\{$GPUs$\}$},
  author={Wang, Yaoyu and Guo, Xiao and Xiao, Junmin and Chen, De and Tan, Guangming},
  booktitle={2025 USENIX Annual Technical Conference (USENIX ATC 25)},
  pages={417--432},
  year={2025}
}

@article{libra,
  title={Libra: Synergizing CUDA and Tensor Cores for High-Performance Sparse Matrix Multiplication},
  author={Shi, Jinliang and Li, Shigang and Xu, Youxuan and Wang, Xueying and Fu, Rongtian and Ma, Zhi and Wu, Tong},
  journal={arXiv e-prints},
  pages={arXiv--2506},
  year={2025}
}

@inproceedings{brp,
  title={BRP-SpMM: Block-Row Partition Based Sparse Matrix Multiplication with Tensor and CUDA Cores},
  author={Dong, Yukang and Jiang, Wenbin and Shen, Xinhai and Guo, Haihong and Shao, Zhiyuan and Jin, Hai},
  booktitle={2025 IEEE International Parallel and Distributed Processing Symposium (IPDPS)},
  pages={901--912},
  year={2025},
  organization={IEEE}
}

@inproceedings{moustafa,
  title={Accelerating sparse matrix-matrix multiplication with the Ascend AI core},
  author={Moustafa, Salli},
  booktitle={Proc. Workshop Accelerated Mach. Learn},
  pages={33--41},
  year={2023}
}

@article{pyg,
  title={PyG 2.0: Scalable learning on real world graphs},
  author={Fey, Matthias and Sunil, Jinu and Nitta, Akihiro and Puri, Rishi and Shah, Manan and Stojanovi{\v{c}}, Bla{\v{z}} and Bendias, Ramona and Barghi, Alexandria and Kocijan, Vid and Zhang, Zecheng and He, Xinwei and Lenssen, Jan Eric and Leskovec, Jure},
  journal={arXiv preprint arXiv:2507.16991},
  year={2025}
}

@manual{mindgl,
  author       = {Mindspore},
  title        = {MindSpore Graph Learning},
  year         = {2022},
  url          = {https://github.com/mindspore-ai/graphlearning},
  note         = {mindgl}
}

@inproceedings{DGL,
  title={Deep graph library: Towards efficient and scalable deep learning on graphs},
  author={Wang, Minjie Yu},
  booktitle={ICLR workshop on representation learning on graphs and manifolds},
  year={2019}
}

@article{suitesparse,
  title={The University of Florida sparse matrix collection},
  author={Davis, Timothy A and Hu, Yifan},
  journal={Acm transactions on mathematical software (toms)},
  volume={38},
  number={1},
  pages={1--25},
  year={2011},
  publisher={ACM New York, NY, USA}
}

@article{yelp,
  title={Graphsaint: Graph sampling based inductive learning method},
  author={Zeng, Hanqing and Zhou, Hongkuan and Srivastava, Ajitesh and Kannan, Rajgopal and Prasanna, Viktor},
  journal={arXiv preprint arXiv:1907.04931},
  year={2019}
}

@article{ogb,
  title={Open graph benchmark: Datasets for machine learning on graphs},
  author={Hu, Weihua and Fey, Matthias and Zitnik, Marinka and Dong, Yuxiao and Ren, Hongyu and Liu, Bowen and Catasta, Michele and Leskovec, Jure},
  journal={Advances in neural information processing systems},
  volume={33},
  pages={22118--22133},
  year={2020}
}

@article{GCN,
  title={Semi-supervised classification with graph convolutional networks},
  author={Kipf, Thomas N and Welling, Max},
  journal={arXiv preprint arXiv:1609.02907},
  year={2016}
}

@article{sci1,
  title={Matrix Krylov subspace methods for linear systems with multiple right-hand sides},
  author={Heyouni, Mohammed and Essai, Azeddine},
  journal={Numerical Algorithms},
  volume={40},
  number={2},
  pages={137--156},
  year={2005},
  publisher={Springer}
}

@inproceedings{sci2,
  title={Accelerating the LOBPCG method on GPUs using a blocked sparse matrix vector product.},
  author={Anzt, Hartwig and Tomov, Stanimire and Dongarra, Jack J},
  booktitle={SpringSim (HPS)},
  pages={75--82},
  year={2015}
}

@inproceedings{llm1,
  title={Sparsegpt: Massive language models can be accurately pruned in one-shot},
  author={Frantar, Elias and Alistarh, Dan},
  booktitle={International conference on machine learning},
  pages={10323--10337},
  year={2023},
  organization={PMLR}
}

@inproceedings{llm2,
  title={A simple and effective pruning approach for large language models},
  author={Sun, Mingjie and Liu, Zhuang and Bair, Anna and Kolter, Zico},
  booktitle={International Conference on Learning Representations},
  volume={2024},
  pages={4942--4964},
  year={2024}
}

@article{llm3,
  title={Flash-llm: Enabling cost-effective and highly-efficient large generative model inference with unstructured sparsity},
  author={Xia, Haojun and Zheng, Zhen and Li, Yuchao and Zhuang, Donglin and Zhou, Zhongzhu and Qiu, Xiafei and Li, Yong and Lin, Wei and Song, Shuaiwen Leon},
  journal={arXiv preprint arXiv:2309.10285},
  year={2023}
}

@inproceedings{llm4,
  title={Spinfer: Leveraging low-level sparsity for efficient large language model inference on gpus},
  author={Fan, Ruibo and Yu, Xiangrui and Dong, Peijie and Li, Zeyu and Gong, Gu and Wang, Qiang and Wang, Wei and Chu, Xiaowen},
  booktitle={Proceedings of the Twentieth European Conference on Computer Systems},
  pages={243--260},
  year={2025}
}

@article{gnnsurvey,
  title={Computing graph neural networks: A survey from algorithms to accelerators},
  author={Abadal, Sergi and Jain, Akshay and Guirado, Robert and L{\'o}pez-Alonso, Jorge and Alarc{\'o}n, Eduard},
  journal={ACM Computing Surveys (CSUR)},
  volume={54},
  number={9},
  pages={1--38},
  year={2021},
  publisher={ACM New York, NY}
}

@inproceedings{GIN,
  title={Net: Degree-specific graph neural networks for node and graph classification},
  author={Wu, Jun and He, Jingrui and Xu, Jiejun},
  booktitle={Proceedings of the 25th ACM SIGKDD international conference on knowledge discovery \& data mining},
  pages={406--415},
  year={2019}
}

@article{GAT,
  title={Graph attention networks},
  author={Veli{\v{c}}kovi{\'c}, Petar and Cucurull, Guillem and Casanova, Arantxa and Romero, Adriana and Lio, Pietro and Bengio, Yoshua},
  journal={arXiv preprint arXiv:1710.10903},
  year={2017}
}

@article{seerattn,
  title={Seerattention: Learning intrinsic sparse attention in your llms},
  author={Gao, Yizhao and Zeng, Zhichen and Du, Dayou and Cao, Shijie and Zhou, Peiyuan and Qi, Jiaxing and Lai, Junjie and So, Hayden Kwok-Hay and Cao, Ting and Yang, Fan and Yang, Mao},
  journal={arXiv preprint arXiv:2410.13276},
  year={2024}
}

@inproceedings{hcspmm,
  title={HC-SpMM: Accelerating Sparse Matrix-Matrix Multiplication for Graphs with Hybrid GPU Cores},
  author={Li, Zhonggen and Ke, Xiangyu and Zhu, Yifan and Gao, Yunjun and Tu, Yaofeng},
  booktitle={2025 IEEE 41st International Conference on Data Engineering (ICDE)},
  pages={501--514},
  year={2025},
  organization={IEEE}
}

@article{neutronsparse,
  title={NeutronSparse: Coordinating Heterogeneous Engines for Sparse Matrix Multiplication on NPUs},
  author={Ai, Xin and Ling, Zeyu and Yuan, Hao and Wang, Qiange and Zhang, Yanfeng and Peng, Yutao and Yu, Ge},
  journal={arXiv preprint arXiv:2606.22482},
  year={2026}
}

\end{document}